\documentclass[10pt]{article}
\usepackage{multicol}
\usepackage{graphicx}
\usepackage{amsmath}
\usepackage[a4paper]{geometry}
\usepackage{hyperref}
\usepackage{rotating}
\usepackage{xcolor}

\begin{document}

\begin{center}

{\Large \bf Collectivity Signatures in High-Multiplicity pp Collisions from Hybrid Hydro+Tsallis Modeling of Pion Spectra}

\vskip1.0cm
Murad Badshah$^{1,}${\footnote{Corresponding author: murad\_phy@awkum.edu.pk; muradbadshah25295@gmail.com}},
Haifa I. Alrebdi$^{2}${\footnote{hialrebdi@pnu.edu.sa}}
Muhammad Waqas$^{3,}${\footnote{Corresponding author: 20220073@huat.edu.cn}}
Hadiqa Qadir$^{1,}${\footnote{awkum-240218506@awkum.edu.pk}}
Muhammad Ajaz$^{1,}${\footnote{Corresponding author: ajaz@awkum.edu.pk}}

{\small\it $^1$Department of Physics, Abdul Wali Khan University Mardan, Mardan 23200, Pakistan \\
$^2$Department of Physics, College of Science, Princess Nourah bint Abdulrahman University, P.O. Box 84428, Riyadh 11671, Saudi Arabia \\
$^3$Hubei Key Laboratory of Energy Storage and Power Battery, School of Optoelectronic Engineering, School of New Energy, Hubei University of Automotive Technology, Shiyan 442002, China\\

}
\end{center}

{\bf Abstract:}
The transverse momentum ($p_T$) distributions up to $p_T = 20$ GeV/c for pions ($\pi^+ + \pi^-$) produced in the ten different multiplicity classes (MCs) of symmetric pp collisions at $\sqrt{s}$ = 7 TeV have been investigated. Two distinct models, the Tsallis-Pareto type function (model) and the combined Boltzmann Gibbs Blast Wave (BGBW) model and Tsallis-Pareto type model (combined Hydro+Tsallis) have been employed to fit the $p_T$ distributions via the minimum $\chi^2$ method. The combined Hydro+Tsallis model is more reliably describing the $p_T$ spectra than the Tsallis-Pareto model. The Tsallis temperature ($T$), non-extensivity parameter ($q$), normalization constant ($N_0$), Kinetic freeze-out temperature ($T_0$), transverse flow velocity ($\beta_T$), and mean transverse momentum ($\langle p_T\rangle$) have been extracted through the fitting procedure via the employed models. The Tsallis-Pareto model gives $T$, $q$, $N_0$ and $\langle p_T\rangle$ while Hydro+Tsallis model gives $T_0$, $\beta_T$, $T$, $q$, $N_0$ and $\langle p_T\rangle$. Incorporating the values of the extracted $T$ and $q$ the thermodynamic quantities and response functions, including energy density ($\varepsilon$), particle density ($n$), entropy density ($s$), pressure ($P$), specific heat at constant volume ($C_V$), squared speed of sound (${c_s}^2$), mean free path ($\lambda$), Knudsen number ($K_n$), isothermal compressibility ($\kappa_T$), and expansion coefficient ($\alpha$) have been calculated at the freeze-out stage. It has been observed that $T$, $\beta_T$, $\langle p_T\rangle$, $N_0$, $\varepsilon$, $n$, $s$, $P$, $C_V$, ${c_s}^2$, and $\alpha$ increase with increasing(decreasing) the charged particles multiplicity density $dN_{ch}/d\eta$(MCs). While $T_0$, $q$, $\lambda$, $K_n$, and $\kappa_T$ decrease with increasing(decreasing) $dN_{ch}/d\eta$(MCs). These systematic variations in the trends of parameters might suggest the gradual transition towards collectivity and thermal equilibration in the high multiplicity pp events, possibly signalling enhanced collective dynamics and partial thermalization in small collision systems.         

\vskip1.0cm

\begin{multicols}{2}

{\section{Introduction}}
Relativistic heavy nuclei collisions generate a highly unstable, short-lived matter called the Quark-Gluon Plasma (QGP). In the QGP state, quarks and gluons interact strongly yet exhibit asymptotic freedom, an unusual behavior \cite{busza2018heavy, das2016propagation, das2017effect, gelis2004photon}. The QGP’s extreme temperature and energy density resemble those of the early universe, making it a central topic of high-energy physics research. Numerous collaborations worldwide are investigating QGP to shed light on the early universe conditions. Typically, symmetric proton–proton (pp) and nucleus–nucleus (A–A) collisions, together with asymmetric proton–nucleus (p–A) collisions, are analyzed to probe the dynamics of nuclear matter \cite{4a, 4b, 4c, 4d}. Although QGP formation is confirmed in heavy-ion collisions, such as symmetric Pb-Pb or Au-Au, possible evidence of QGP-like behavior has also been observed in symmetric proton-proton (pp) collisions \cite{strange, sahoo2021possible}, which often serve as baselines for heavy-ion studies. Observable QGP signatures include phenomena like strangeness enhancement, $J/\psi$ suppression or melting, and jet quenching \cite{5, 6, 8}.

Recent data from symmetric pp and anti-symmetric proton-lead (p-Pb) collisions have revealed trends similar to those in symmetric Pb-Pb collisions. Analysis of the $p_T$ distribution in pp collisions demonstrates a clear spectral hardening from low-multiplicity (class-X) to high-multiplicity (class-I) events, paralleling trends in larger collision systems \cite{identified}. Furthermore, studies show that, unlike for pions, the charged-particle multiplicity in these events increases with the strange and multi-strange hadrons integrated yields \cite{strange}, consistent with previously observed strangeness enhancement in Pb-Pb collisions.

The QGP acts like a hot, expanding fluid that cools through various stages, each defined by a distinct temperature. The chemical freeze-out temperature corresponds to the chemical freeze-out stage, where inelastic interactions between fireball constituents cease, halting the production of new particles. As the fireball expands, inter-particle distances increase to the point that inelastic collisions are no longer possible \cite{9, 10, 11}. At the thermal freeze-out stage, elastic collisions between particles also cease, marking the thermal freeze-out temperature \cite{12, 13, 14}. After thermal freeze-out, particles propagate to the detectors with energy and momentum distributions that directly reflect the conditions at this stage. Thus, by analyzing the $p_T$ distributions of the detected particles, we make accurate insights into the kinetic freeze-out conditions, including the kinetic freeze-out temperature, the radial or transverse flow velocity, the system’s post-collision volume (V), the particle multiplicity ($N_0$) and some other crucial thermodynamic quantities. Charged pions are among the most abundantly produced hadrons in high-energy collisions and serve as the primary decay products of resonances and other particles. Consequently, this study focuses on analyzing the $p_T$ distribution of charged pions. 

The detectors are unable to measure the freeze-out parameters mentioned above directly. Consequently, various hydrodynamical and statistical models are employed to estimate these important parameters, contributing to our understanding of the true nature of the QGP. Even after particle interactions cease during the system’s evolution, particles continue to occupy space according to a statistical distribution \cite{15}. In this section, we will briefly review the models used in the present analysis, as well as those that are commonly applied. Non-extensive statistical distributions are widely used and convenient for analyzing the $p_T$ spectra of the generated hadrons in high-energy collisions. Various Tsallis distributions have been employed in the literature to successfully describe the $p_T$ spectra of generated hadrons in pp collisions at RHIC and LHC energies \cite{16, 17, 18, 19, 20, 21}. 

The appeal of the Tsallis distribution lies in its simplicity; it relies only on two free parameters that assist in fitting the data: the Tsallis temperature ($T$), which encompasses both the thermal and flow effects of the system; the non-extensive parameter ($q$), which quantifies disagreement to traditional Boltzmann-Gibbs statistics; and the normalization constant ($N_0$). While traditionally associated with heavy-ion collisions, recent studies have revealed hints of collective behaviour (e.g., azimuthal anisotropies and long-range correlations) in pp collisions at the LHC, motivating further analysis of thermodynamic properties.. These parameters ($q$ and $T$) can be utilized to calculate the thermodynamic quantities and response functions, including energy density ($\varepsilon$), particle density ($n$), entropy density ($s$), pressure ($P$), specific heat at constant volume ($C_V$), squared speed of sound (${c_s}^2$), mean free path ($\lambda$), Knudsen number ($K_n$), isothermal compressibility ($\kappa_T$), and expansion coefficient ($\alpha$).  

To examine the $p_T$ spectra of hadrons emitted in relativistic nucleus collisions, various models with flow effects are incorporated to extract parameters like transverse flow velocity and kinetic freeze-out temperature, models commonly used include the Boltzmann Gibbs Blast Wave (BGBW) model \cite{29, 30}, the Blast-Wave model with Tsallis statistics (TBW model) \cite{27, 28}, the Hagedorn model with embedded transverse flow \cite{23, 24, 25, 26, 26a}, and the Tsallis distribution with flow effects, also known as the improved Tsallis distribution \cite{31, 32, haifa, 33}.
\\
{\section{The method and formalism}} 
The $p_T$ distribution of the generated hadrons in the relativistic collisions carries critical insights of the freeze-out stage of the created system. Thus, the $p_T$ spectrum serves as a fundamental observable for investigating the true nature of the QGP and its different stages of evolution. Historically, the Boltzmann–Gibbs exponential function, defined by Eq. (1), has been extensively applied to describe the $p_T$ spectra of emitted particles:
\begin{align}
f(E, T, \mu) \propto \exp \left(-\frac{E-\mu}{T}\right),
\end{align}
\noindent where \( E \) is the particle's energy, \( \mu \) is the chemical potential, and \( T \) represents the temperature. However, this approach becomes inadequate at high $p_T$ values (specifically, $p_T > 3$ GeV/c), where particle production mechanisms are dominated by non-thermal, non-equilibrium processes associated with perturbative QCD. In this high-$p_T$ regime, a power-law distribution is more effective \cite{waqas2021effects}.

To accommodate this, the Boltzmann–Gibbs framework has been generalized within non-extensive statistical mechanics, yielding the Tsallis–Pareto (Tsallis) distribution function \cite{biro2020tsallis, su2021non, minelatest}. This generalized function for describing particle $p_T$ spectra is expressed in Eq. (2) or Eq. (3):
\begin{align}
f(E, q, T, \mu) = \left[1 + \frac{q - 1}{T} (E - \mu)\right]^{-\frac{1}{q - 1}}.
\end{align}
\noindent For mid-rapidity, where \(\mu = m_0\) (with \(m_0\) as the mass of the generated hadrons) and \(E = m_T = \sqrt{p_T^2 + m_0^2}\), we can rewrite this expression as:
\begin{align}
f\left(m_T, T, q\right) = N_0 \, m_T \left[1 + \frac{q - 1}{T}(m_T - m_0)\right]^{-\frac{q}{q - 1}},
\end{align}
\noindent where \( T \) represents the Tsallis temperature, typically defined as \( T = T_0 + m_0 \langle \beta_T \rangle \), with \( T_0 \) as the kinetic freeze-out temperature and \( \langle \beta_T \rangle \) the mean transverse flow velocity. Here, \( q \) is the non-extensive parameter, which approaches unity as the system nears thermal equilibrium. When \( q = 1 \), the Tsallis–Pareto function converges to the Boltzmann–Gibbs exponential function. \( N_0 \) is the normalization constant, given by \( N_0 = \frac{C_y g V}{(2 \pi)^3} \), where \( C_y \) represents the correction factor specific to the experimental data, \( g \) denotes the degeneracy factor, and \( V \) is the kinetic freeze-out volume of the emitted particles. \( m_T = \sqrt{p_T^2 + m_0^2} \) is the transverse mass, and \( m_0 \) is the particle's rest mass. For pseudo-rapidity differential yields, \( C_y = \frac{m_T}{p_T} \), while for rapidity differential yields, \( C_y = 1 \). This study focuses on rapidity yields as utilized in Eq. (3).

It is recognized that the Tsallis distribution can not fully capture the observed structure in the low-$p_T$ region \cite{nath2021centrality}. This discrepancy highlights the importance of distinguishing between the production mechanisms for soft and hard particles. Improving the pions fit at very small $p_T$ is necessary. The majority of low-$p_T$ particles originate from the QGP produced in high-energy collisions, where they follow an exponential distribution. In contrast, high-$p_T$ particles result from mini jets, which traverse the hot medium and are described by perturbative Quantum Chromodynamics (pQCD). In high-energy colliding systems, collective motion effects also need to be considered. To accurately describe the $p_T$ distribution in a low $p_T$ regime and to incorporate the flow effects the BGBW model \cite{29, 30}, given in Eq. (4), is used. Aiming to improve the fit quality in the small $p_T$ region and the overall fit quality, we combine the BGBW and Tsallis models via the superposition principle, given in Eq. (5). The hybrid model, Eq. (5), is formulated to account for both soft (BGBW) and hard (Tsallis) particle production mechanisms, enabling a more robust description of the full $p_T$ spectrum. This approach yields superior fit quality compared to individual models, as demonstrated later. Though in the references \cite{badshah2024centrality, badshah2024thermodynamic, waqas2020initial, li2018energy} the superposition of two or three components of a single model has been successfully applied to the $p_T$ distribution of the produced hadrons, but here we apply the superposition to combine two different models, i.e., BGBW and Tsallis models.
\begin{align}
f\left(p_T, T_0, \beta_{T}\right)= & N_0 p_T m_T \int_0^R r \mathrm{~d} r\nonumber \\ 
& \times I_0\left[\frac{p_T \sinh \left(\rho_1\right)}{T_0}\right] K_1\left[\frac{m_T \cosh \left(\rho_1\right)}{T_0}\right],
\end{align}
where $I_0$ and $K_1$ are the modified Bessel's first and second kind functions, respectively. The boost angle, $\rho=\tanh ^{-1}\left[\beta_S\left(\frac{r}{R}\right)^{n_0}\right]$, increases proportionally to the $n^{th}$ power of the radial distance ($r$) within the transverse plane, with $\beta_S$ indicating the transverse flow velocity at the surface of the fireball (where $r = R$ and $R$ is the fireball's radius or the maximum value of $r$). The mean transverse flow velocity is given by $\langle \beta_T \rangle = \frac{2}{n_0 + 2} \beta_S$. In our model, default value for $n_0$ is set at 1, resulting in $\langle \beta_T \rangle = \frac{2}{3} \beta_S$.
\begin{align}
f\left(p_T, T_0, \beta_{T}, m_T, T, q\right)=kf\left(p_T, T_0, \beta_{T}\right)+\nonumber \\
(1-k)f\left(m_T, T, q\right),
\end{align}
where $k$ is the contribution fraction of the BGBW model to the Tsallis model. The superposition of the two models given in Eq. (5) is referred to as the BGBW+Tsallis model or the Hydro+Tsallis model throughout the manuscript. It is noteworthy that the parameters of the BGBW model and Tsallis model in the combined Hydro+Tsallis model are treated as uncorrelated. This treatment allows for greater flexibility to capture soft (hydro-like) and hard (non-thermal) components of the $p_T$ spectra. 

Regardless of the specific structure of the particles $p_T$ spectra, the probability density function (PDF) as a function of $p_T$ is defined by:
\begin{align}
f(p_T) = \frac{1}{N}\frac{dN}{dp_T}, 
\end{align}
which ensures normalization to unity:
\begin{align}
\int_{0}^{\infty} f(p_T)\ dp_T = 1. 
\end{align}
Here $f(p_T)$ means the Tsallis and Hydro+Tsallis distribution functions. The mean $p_T$, $\langle p_T \rangle$, can be derived directly from the fit function using the PDF as follows:
\begin{align}
\langle p_T \rangle = \frac{\int_{0}^{\infty} p_T\ f(p_T)\ dp_T}{\int_{0}^{\infty} f(p_T)\ dp_T},
\end{align}
substituting Eq. (7) into this expression simplifies to:
\begin{align}
\langle p_T \rangle = \int_{0}^{\infty} p_T\ f(p_T)\ dp_T. 
\end{align}

The values of $T$ and $q$ obtained from Tsallis distribution are used to calculate very crucial thermodynamic quantities at the freeze-out stage \cite{wu2023thermodynamic, jain2023study, badshah2024dependence}
\begin{align}
\varepsilon=  g \int \frac{\mathrm{d}^3 p}{(2 \pi)^3} E\left[1+(q-1) \frac{E}{T}\right]^{\frac{q}{1-q}},
\end{align}
\begin{align}
n=  g \int \frac{\mathrm{d}^3 p}{(2 \pi)^3}\left[1+(q-1) \frac{E}{T}\right]^{\frac{q}{1-q}},
\end{align}
\begin{align}
s & = g \int \frac{\mathrm{d}^3 p}{(2 \pi)^3}\left[\frac{E}{T}\left(1+(q-1) \frac{E}{T}\right)^{\frac{q}{1-q}}\right. \nonumber \\
& \left.+\left(1+(q-1) \frac{E}{T}\right)^{\frac{1}{1-q}}\right],
\end{align}
\begin{align}
P=  g \int \frac{\mathrm{d}^3 p}{(2 \pi)^3} \frac{p^2}{3 E}\left[1+(q-1) \frac{E}{T}\right]^{\frac{q}{1-q}},
\end{align}
In these equations, $g$ is the degeneracy factor, and $E$ is the energy of the produced particles.

Another important group of quantities, although not directly observable, plays an important role in understanding the characteristic of the medium and the equation of state: the thermodynamic response functions. These include variables that describe how a system reacts to changes in external parameters such as pressure and temperature. The response functions include the heat capacity at constant volume $C_V$, squared speed of sound $c_s^2$ \cite{wu2023thermodynamic, jain2023study}, mean free path $\lambda$ \cite{sahu2021characterizing}, Knudsen number ($K_n$) \cite{sarkar2019finite}, isothermal compressibility ($\kappa_T$) \cite{khuntia2019effect}, and expansion coefficient ($\alpha$) \cite{sahu2021characterizing}.
\begin{align}
C_{\mathrm{V}}= g \frac{q}{T^2} \int \frac{\mathrm{d}^3 p}{(2 \pi)^3} {E}^2\left[1+(q-1) \frac{E}{T}\right]^{\frac{2 q-1}{1-q}},
\end{align}
\begin{align}
c_{\mathrm{s}}^2(T)=\left(\frac{\partial P}{\partial \varepsilon}\right)_{\mathrm{V}}=\frac{s}{C_{\mathrm{V}}}.
\end{align}
$\lambda$ is the mean distance a particle traverse between two consecutive collisions within the system. Mathematically, it is given by:
\begin{align}
\lambda = \frac{1}{n\sigma},
\end{align}
where \( n \) denotes the particle density and \( \sigma \) represents the scattering cross-section. In the specific context of the hard-core hadron radius approximation, where \( r_h = 0.3 \) fm, the cross-section is expressed as \( \sigma = 4\pi r_h^2 \) \cite{kadam2015dissipative} Additionally, the Knudsen number is a dimensionless parameter defined as the ratio of the $\lambda$ to the radius \( R \) of the hadronic system, where \( R = \left(\frac{3V}{4\pi}\right)^{1/3} \) \cite{sarkar2019finite}, where $V$ is the produced system's volume at kinetic freeze-out. A small Knudsen number indicates that the system is in the continuum regime, enabling the use of hydrodynamic models. Typically, when \( Kn < 1 \), the criteria for fluid dynamics are met, signifying fluid-like behaviour. Conversely, values exceeding unity suggest that the particles predominantly undergo free-streaming. The isothermal compressibility $\kappa_T$ and expansion coefficient $\alpha$ respectively are given as
\begin{align}
\kappa_{\mathrm{T}}=\frac{\partial n / \partial \mu}{n^2},
\end{align}
where
\begin{align}
\frac{\partial n}{\partial \mu}=\frac{g q}{T} \int \frac{d^3 p}{(2 \pi)^3}\left[1+(q-1) \frac{E}{T}\right]^{\frac{1-2 q}{q-1}},
\end{align}

The (volumetric) thermal expansion coefficient (also known as the expansivity) can be defined as
\begin{equation}
\alpha \;\equiv\; \frac{1}{V}\left(\frac{\partial V}{\partial T}\right)_{p},
\label{eq:alpha_def}
\end{equation}
i.e., the change in volume as a fraction of the change in temperature at constant pressure.
Since the kinetic freeze-out volume for the identified hadrons is not accurately measurable, we use $N/V=n$ and write equivalently \cite{sahu2021characterizing}
\begin{equation}
\alpha \;=\; -\frac{1}{n}\left(\frac{\partial n}{\partial T}\right)_{p}.
\label{eq:alpha_def_density}
\end{equation}
Now making use of Eq. (11), we finally obtain
\begin{align}
\alpha=-\frac{\int d^3 p \left(\frac{E}{T^2}\right) q\left[1+(q-1) \frac{E}{T}\right]^{\frac{1-2 q}{q-1}}}{\int d^3 p\left[1+(q-1) \frac{E}{T}\right]^{\frac{-q}{q-1}}}.
\end{align}

{\section{Results and discussion}}
Fig. 1 represents the double differential $p_T$ distribution of $\pi^+ + \pi^-$ recorded by ALICE at LHC in different MCs of pp collisions \cite{identified}. The different geometrical shapes are used for the experimental data while solid lines are the fitting results of the Tsallis-Pareto model (Fig. 1(a)) and Hydro+Tsallis model (Fig. 1(b)). In order to create improved visualization and to avoid the overlap of the data of various multiplicity classes (MCs) in Fig. 1, the data and the corresponding fit lines have been multiplied by different scaling factors i.e., the data have been scaled. The scaling factors of $10^4, 10^3, ..... 10^{-5}$ have been used for MCs I, II, ..... X, respectively. The minimum $\chi^2$ method has been employed for the fitting procedure, which is given as \cite{bashir2015centrality}
\begin{align}
\chi^2=\sum_{i}\frac{(R_i^{Exp}-R_i^{Theor})^2}{\sigma^2}.
\end{align}
In this case, \(R_i^{Exp}\) is the experimental value of the spectrum in the i-th $p_T$ bin (i.e. the ALICE data point), \(R_i^{Theor}\) is the model predicted value of the same $p_T,i$ (i.e. the bin center), and \(\sigma\) is the uncertainty of the i-th data point (the point-by-point uncertainty used as the weight in the fit). We also determine the number of degrees of freedom as $NDF = N - n_{par}$, where $N$ and $n_{par}$ are the number of experimental data points and number of free fit parameters, respectively, and report $\chi^2/NDF$ in Fig. 2(a).
\begin{figure*}
\centering
\includegraphics[width=0.49\textwidth]{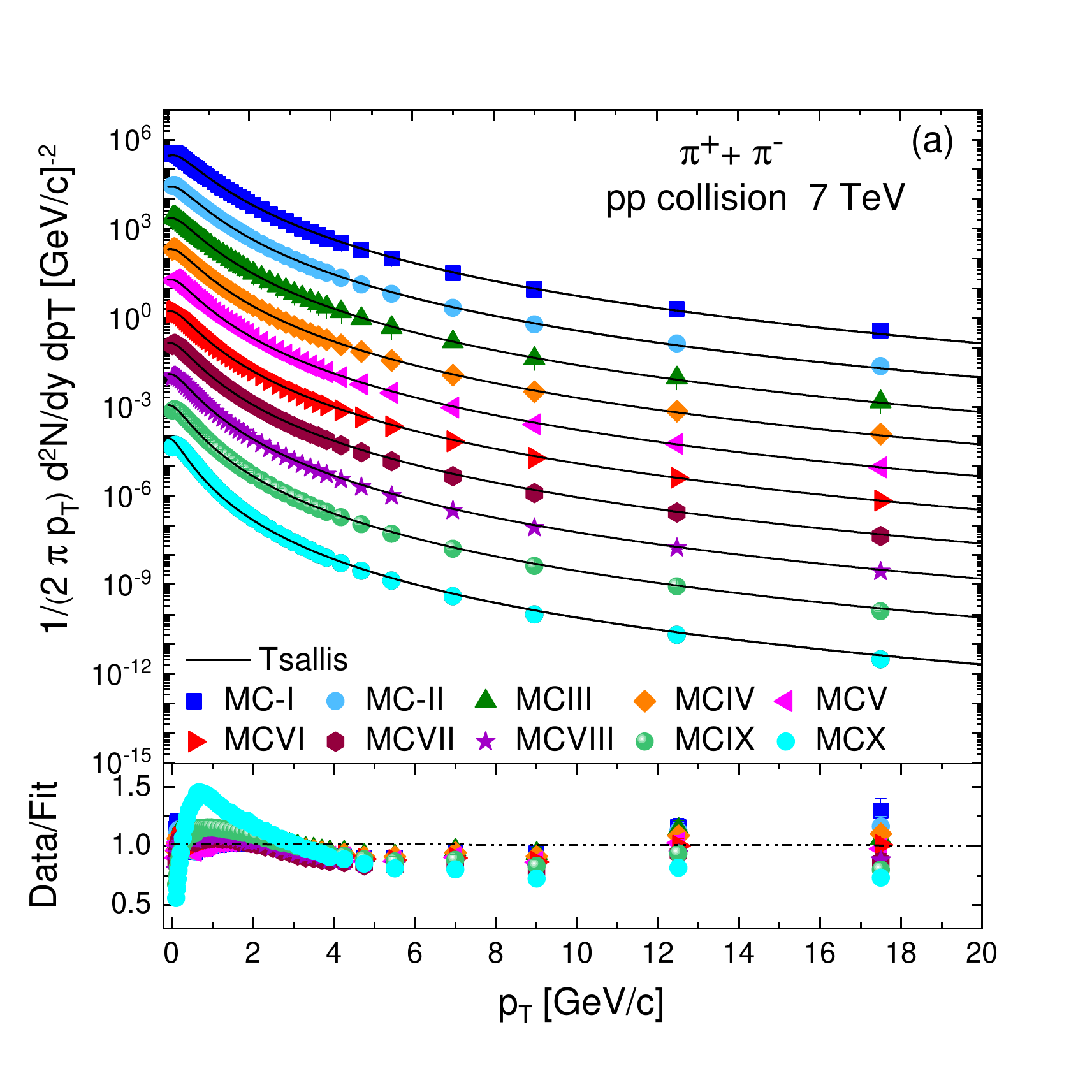}
\includegraphics[width=0.49\textwidth]{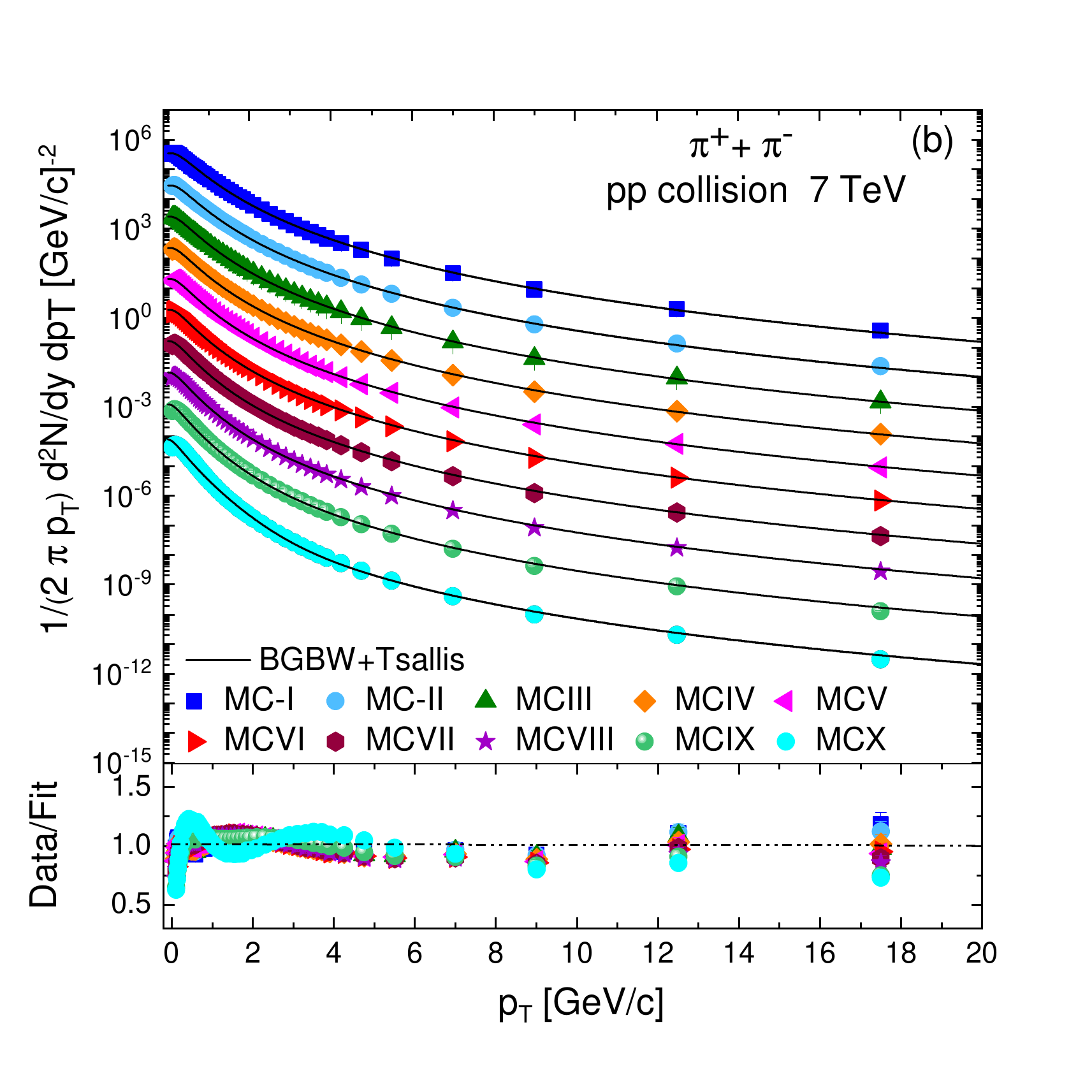}
\caption {Double differential $p_T$ distribution of charged pions produced in pp collisions at $\sqrt{s}$=7 TeV in different multiplicity classes. The various geometrical shapes and colours are used for the experimental data \cite{identified} while the solid lines are used for (a) the Tsallis model and (b) the Hydro+Tsallis model.}
\end{figure*}

It is noteworthy that the parameter $\chi^2$ does not actually determine a universal threshold model preference, especially when the models under competition differ in the number of parameters. Thus, we compared the two descriptions using criteria based on information that is sensitive to model complexity besides $\chi^2/NDF$. Based on the values of $\chi^2$ in Tables 1 and 2, the number of free parameters (3 in Tsallis, and 6 in Hydro+Tsallis) in the models used, and the total number of data points (48). We find that Hydro+Tsallis has a lower $\chi^2/NDF$ in 9 of 10 multiplicity classes (the only exception is MCV, where the difference is relatively small). High multiplicity classes are the most improved; e.g., in MCX, $\chi^2/NDF$ declines from 629.6968/45 to 198.1003/42. At the expense of extra parameters, Hydro+Tsallis is preferred by Akaike Information Criterion ($AIC$) in 8/10 classes and Bayesian Information Criterion ($BIC$) in 7/10 classes, and the summed $\Delta AIC$ and $\Delta BIC$ over classes are highly negative, showing that the enhanced description (Hydro+Tsallis model) is not just due to increased flexibility, i.e, not due to increased number of free parameters. Here $\Delta AIC=AIC_H-AIC_T$ (similarly for $\Delta BIC$), where $AIC_H$ and $AIC_T$ stand for Hydro+Tsallis and pure Tsallis functions, respectively. $\chi^2/NDF$, $AIC$, and $BIC$ as a function of MCs are shown in Fig. 2.

\begin{figure*}
\centering
\includegraphics[width=0.49\textwidth]{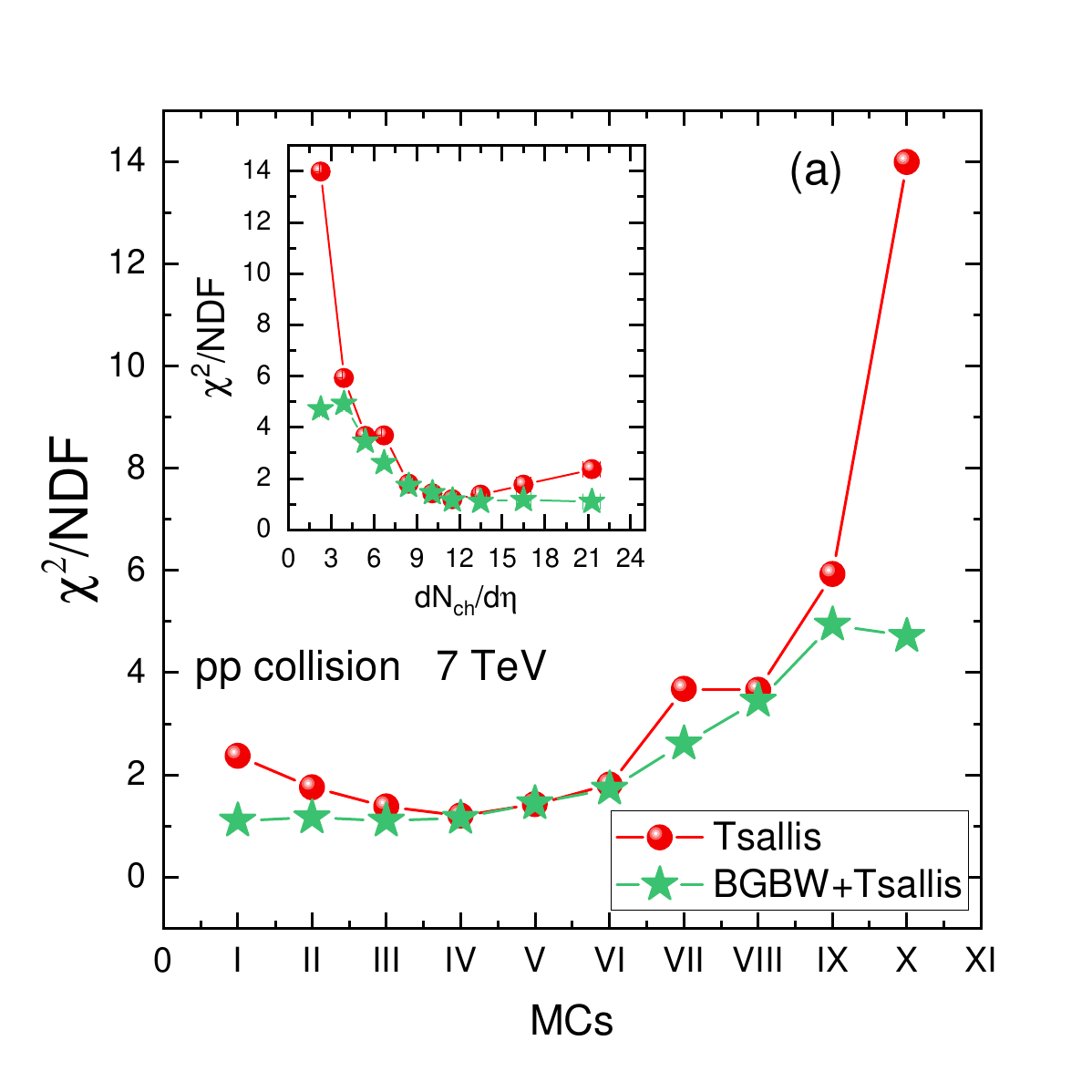}
\includegraphics[width=0.49\textwidth]{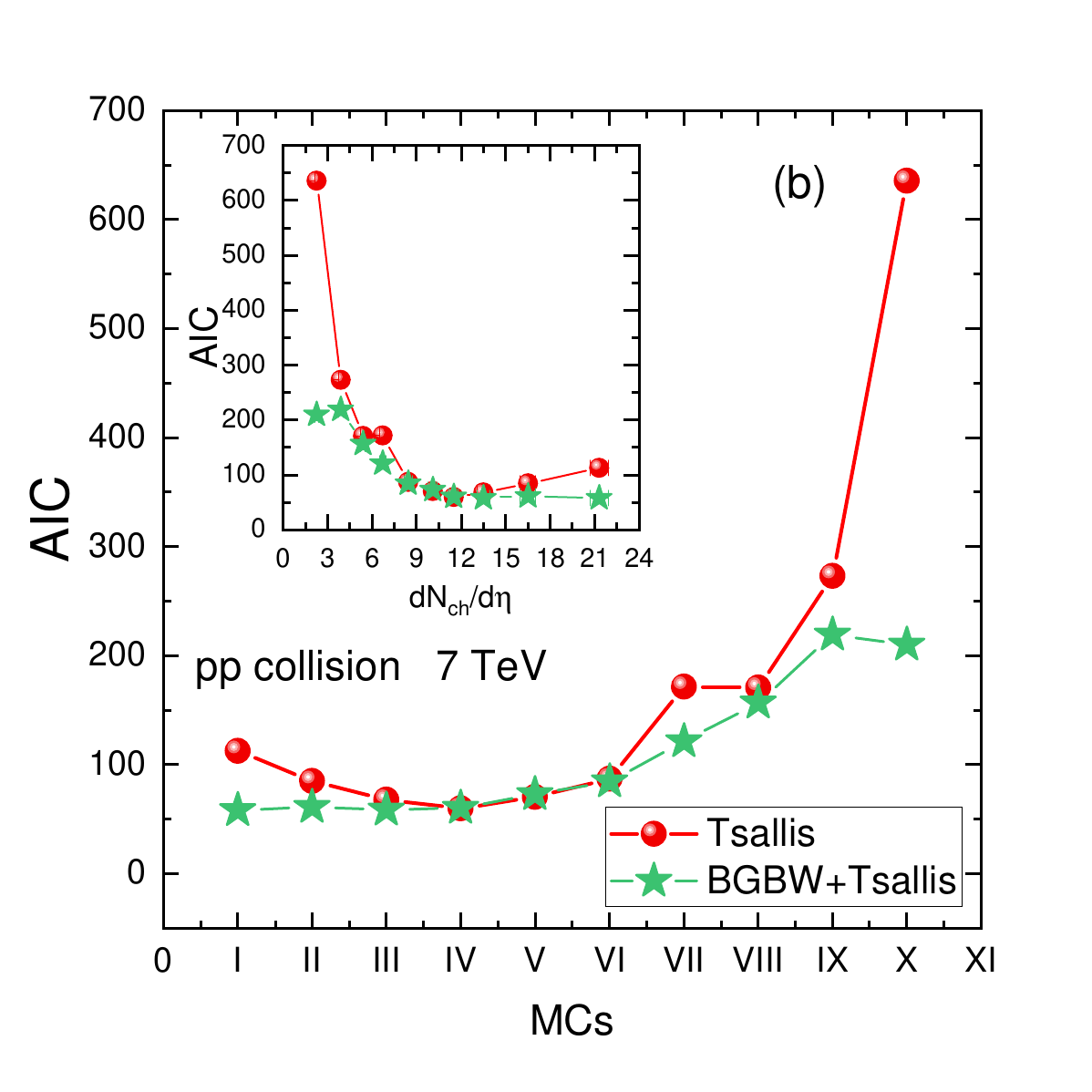}
\includegraphics[width=0.49\textwidth]{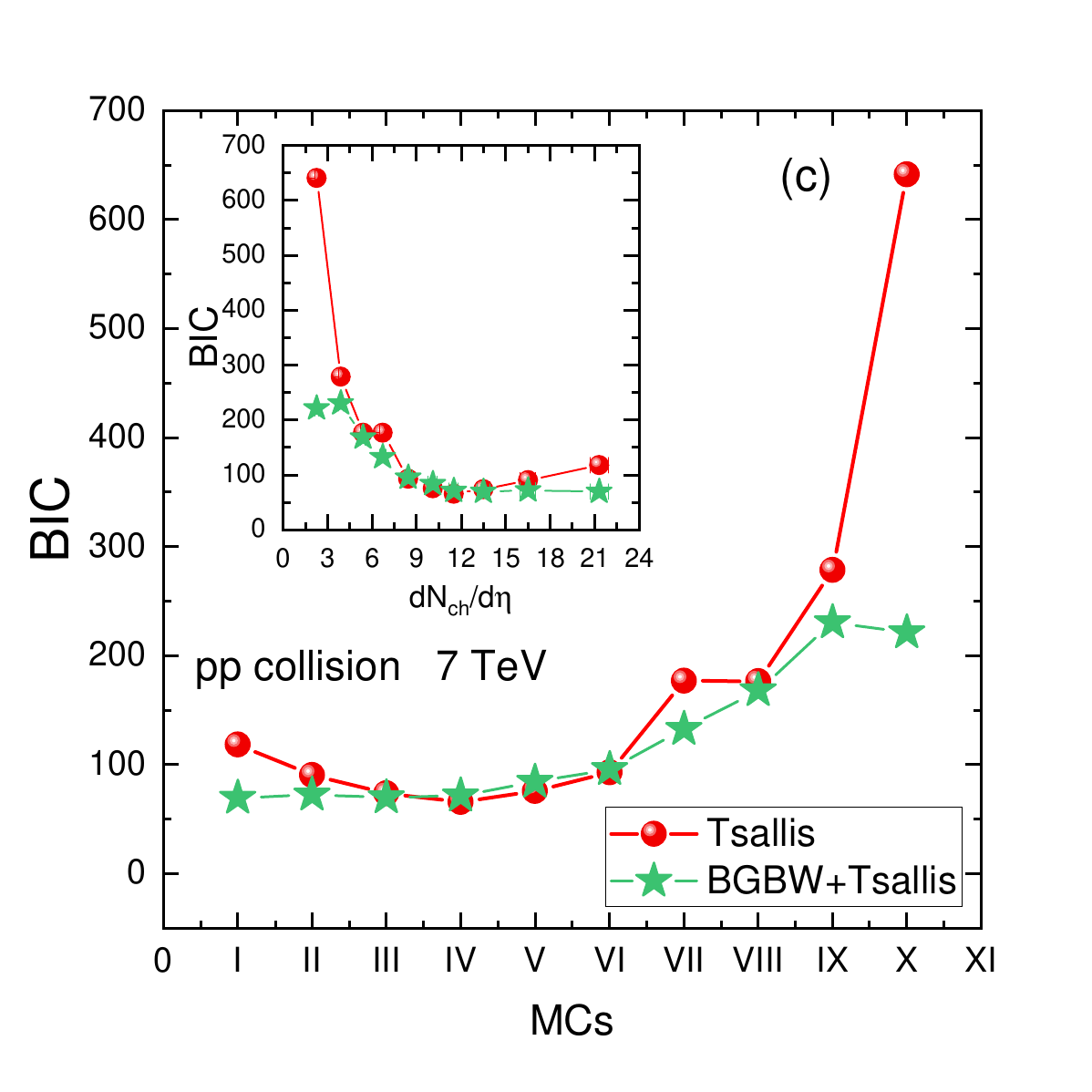}
\caption {(a) $\chi^2/NDF$ (reduced-$\chi^2$), (b) Akaike Information Criterion ($AIC$), and (c) Bayesian Information Criterion ($BIC$) as a function of MCs and $dN_{ch}/d\eta$.}
\end{figure*}
\begin{table*}[!ht]
    \centering
    \caption{The values of different parameters extracted from the Tsallis model along with the values of $\langle p_T\rangle$ and $\chi^2/NDF$.}
    \begin{tabular}{ccccccc}
    \hline
    \hline
        collision & MC & $T$ [GeV] & $q$ & $N_0$ & $\langle p_T\rangle$ [GeV/c] & $\chi^2/NDF$ \\ 
        \hline
    \hline
        \textbf{} & MC1 & 0.2079±0.003 & 1.171±0.0028 & 2024±4.479 & 0.5286±0.0367 & 106.7039/45 \\ 
        \textbf{} & MC2 & 0.1949±0.0028 & 1.174±0.0029 & 1664±4.7107 & 0.4872±0.0351 & 79.0180/45 \\ 
        \textbf{} & MC3 & 0.1861±0.0029 & 1.1751±0.0029 & 1364±4.4626 & 0.463±0.0394 & 62.0456/45 \\ 
               pp & MC4 & 0.1802±0.0031 & 1.1762±0.0031 & 1194±5.7423 & 0.4472±0.0362 & 53.9420/45 \\ 
            7 TeV & MC5 & 0.1752±0.0026 & 1.1771±0.0032 & 1094±5.1446 & 0.4337±0.029 & 64.2184/45 \\ 
        Fig. 1(a) & MC6 & 0.1702±0.003 & 1.1781±0.0032 & 894±4.809 & 0.4203±0.0372 & 81.2392/45 \\ 
        \textbf{} & MC7 & 0.1642±0.0029 & 1.1793±0.0032 & 728±4.9304 & 0.4043±0.03131 & 165.5815/45 \\ 
        \textbf{} & MC8 & 0.1554±0.0029 & 1.1796±0.0029 & 608±5.0271 & 0.3795±0.0304 & 164.8919/45 \\ 
        \textbf{} & MC9 & 0.1401±0.0033 & 1.1806±0.0031 & 478±4.6589 & 0.3372±0.0323 & 266.9641/45 \\ 
        \textbf{} & MC10 & 0.1171±0.0027 & 1.1813±0.0025 & 298±4.7127 & 0.2745±0.0277 & 629.6968/45 \\ 
        \hline
    \hline
    \end{tabular}
\end{table*}
\begin{sidewaystable*}
    \centering
    \caption{The values of different parameters extracted from the Hydro+Tsallis model along with the $\langle p_T\rangle$ and $\chi^2/NDF$ values.}
    \begin{tabular}{cccccccccc}
     \hline
        \hline
        collision & MC &  $T_0$ [GeV] & $\beta_T$ [c] & K & $T$ [GeV] & q & $N_0$ & $\langle p_T\rangle$ [GeV/c] & $\chi^2/NDF$ \\ 
        \hline
        \hline
        \textbf{} & MC1 & 0.1351±0.004 & 0.6201±0.005 & 0.08±0.005 & 0.2±0.0032 & 1.1765±0.0029 & 2192±4.7364 & 0.5051±0.0355 & 46.3524/42 \\ 
        \textbf{} & MC2 & 0.1381±0.0041 & 0.6061±0.005 & 0.08±0.0052 & 0.191±0.0033 & 1.1774±0.0032 & 1722±4.178 & 0.4798±0.0307 & 49.1013/42 \\ 
        \textbf{} & MC3 & 0.1393±0.004 & 0.5963±0.0049 & 0.07±0.0049 & 0.183±0.0027 & 1.1783±0.003 & 1442±4.8534 & 0.4576±0.0351 & 46.5490/42 \\ 
        pp & MC4 & 0.1397±0.0041 & 0.5896±0.0049 & 0.07±0.0051 & 0.178±0.0031 & 1.1793±0.003 & 1241±5.1367 & 0.4441±0.0366 & 48.4990/42 \\ 
        7 TeV & MC5 & 0.1405±0.004 & 0.5796±0.005 & 0.085±0.0052 & 0.1738±0.0031 & 1.1802±0.003 & 1091±4.8511 & 0.4329±0.0336 & 60.7692/42 \\ 
        Fig. 1(b) & MC6 & 0.1424±0.0039 & 0.5658±0.0052 & 0.108±0.005 & 0.1685±0.0029 & 1.181±0.0027 & 945±5.0165 & 0.4184±0.0342 & 72.6387/42 \\ 
        \textbf{} & MC7 & 0.1435±0.004 & 0.5558±0.0049 & 0.118±0.0055 & 0.1605±0.0027 & 1.1818±0.0029 & 775±4.7795 & 0.396±0.0317 & 109.6006/42 \\ 
        \textbf{} & MC8 & 0.1465±0.0041 & 0.5426±0.0047 & 0.206±0.0072 & 0.1566±0.003 & 1.1824±0.003 & 645±4.2817 & 0.3855±0.028 & 145.0399/42 \\ 
        \textbf{} & MC9 & 0.1474±0.0041 & 0.5306±0.0047 & 0.386±0.0081 & 0.1506±0.0028 & 1.1832±0.0032 & 495±4.7053 & 0.369±0.0301 & 207.3622/42 \\ 
        \textbf{} & MC10 & 0.1478±0.004 & 0.5266±0.0046 & 0.736±0.0101 & 0.1483±0.0031 & 1.1841±0.0027 & 290±5.0776 & 0.3632±0.0331 & 198.1003/42 \\
         \hline
        \hline
    \end{tabular}
\end{sidewaystable*}

\begin{figure*}
\centering
\includegraphics[width=0.49\textwidth]{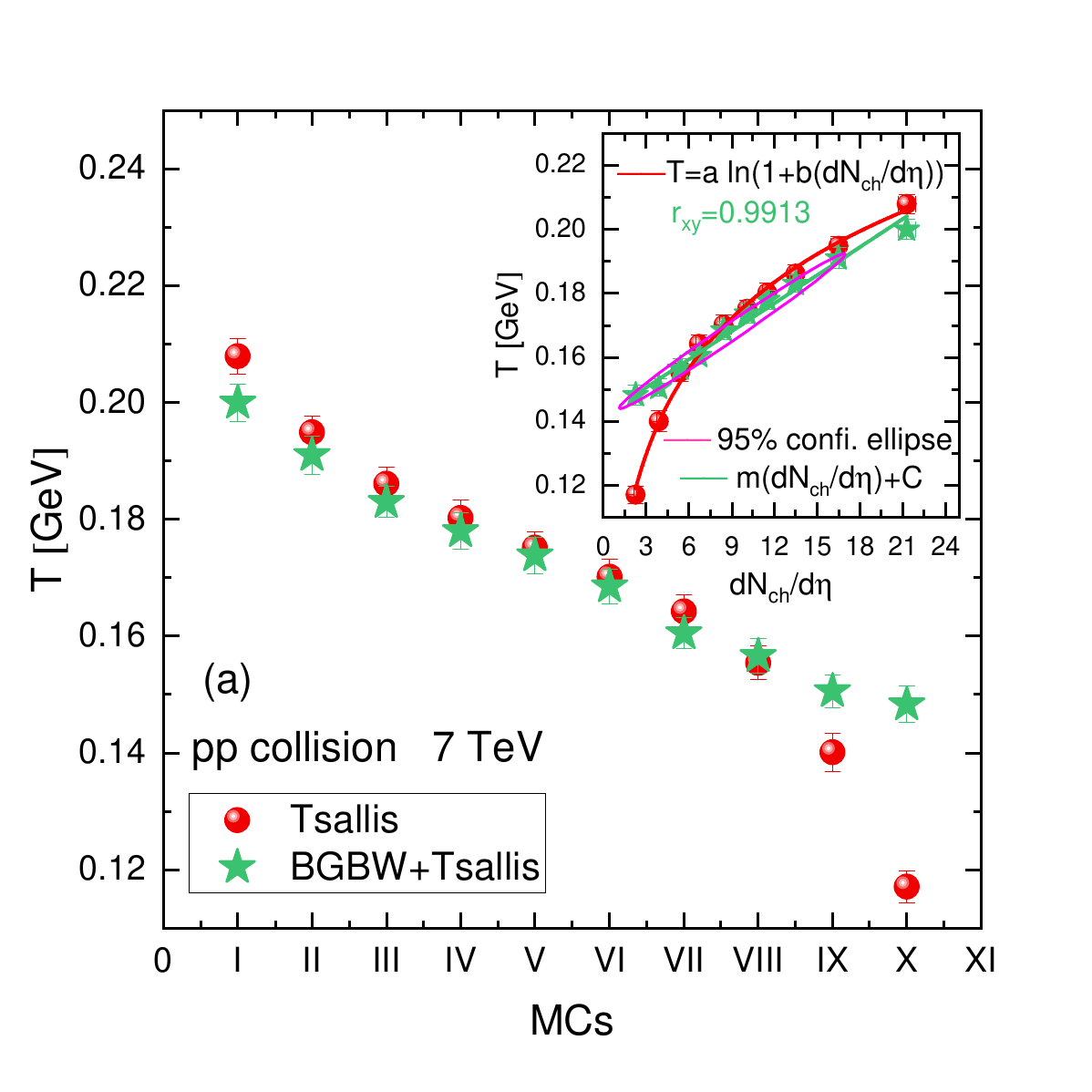}
\includegraphics[width=0.49\textwidth]{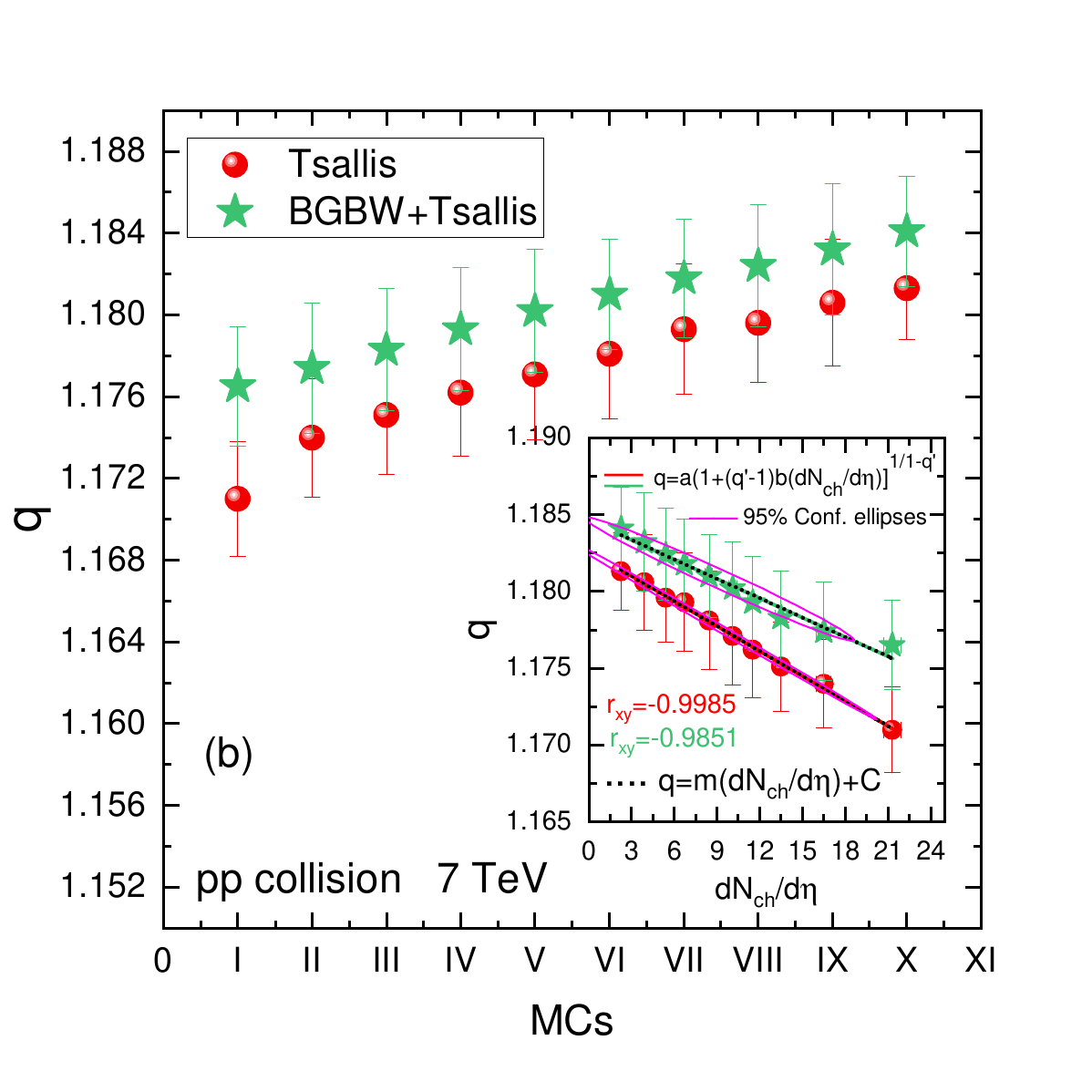}\vspace{-0.35cm}
\includegraphics[width=0.49\textwidth]{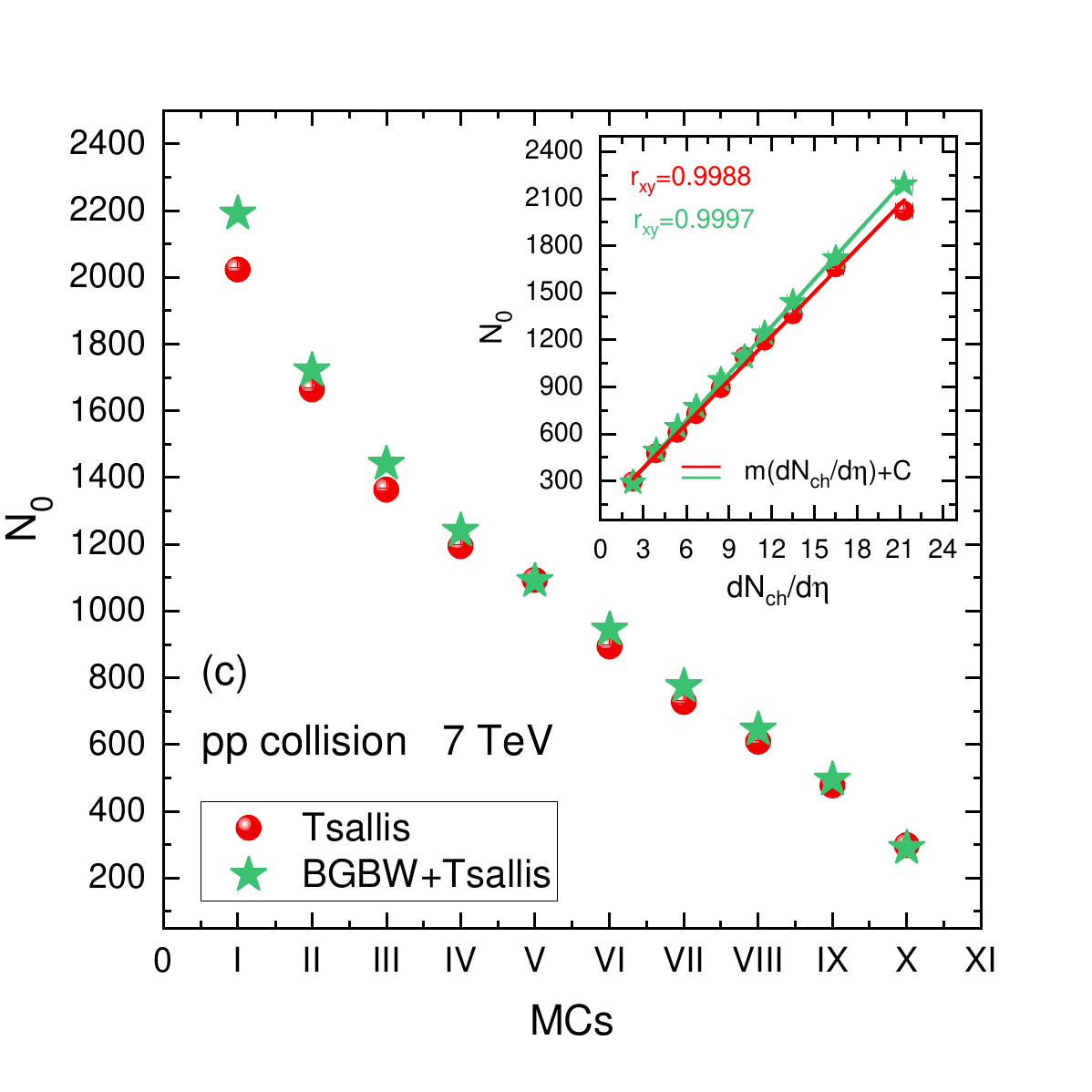}
\includegraphics[width=0.49\textwidth]{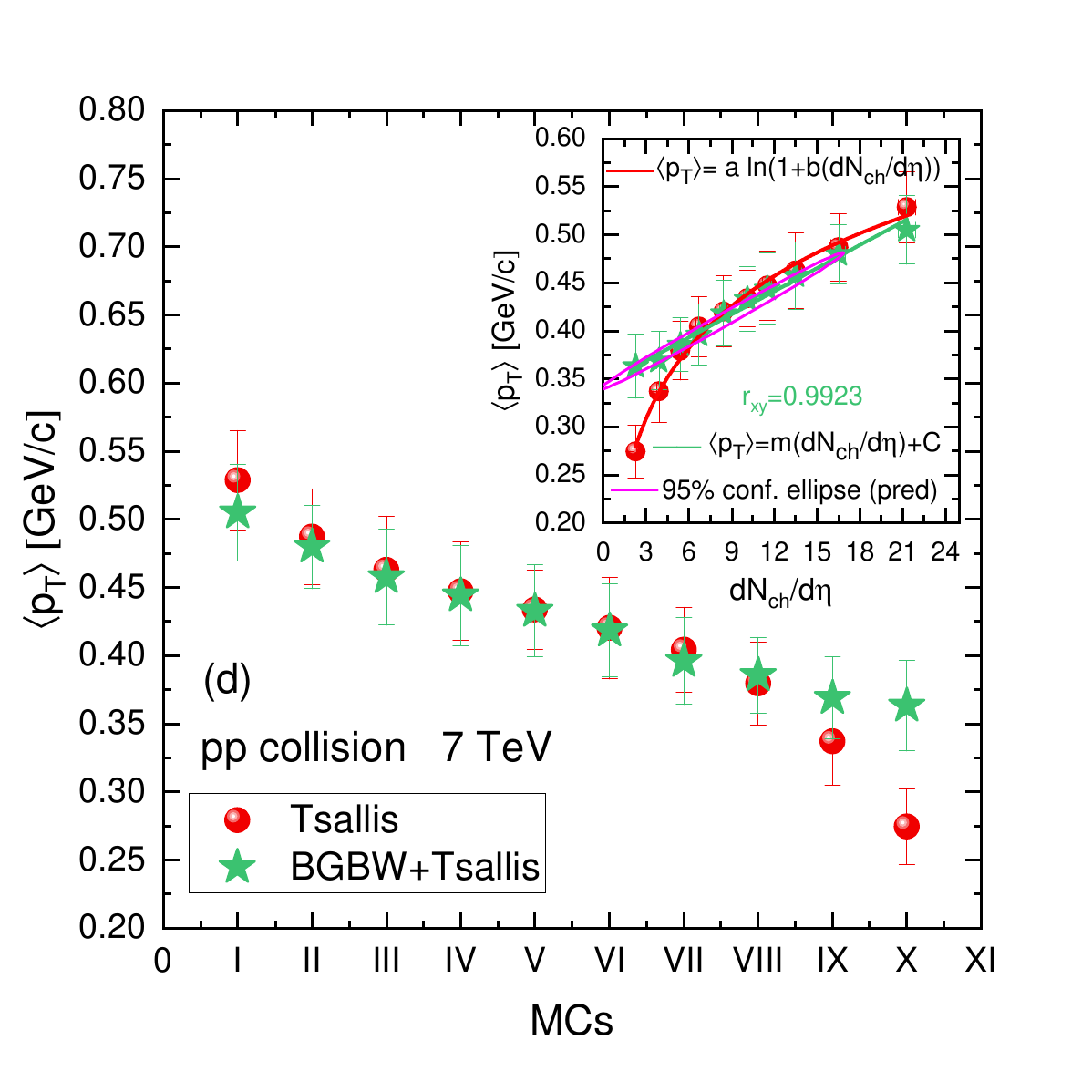}\vspace{-0.35cm}
\caption {(a) $T$, (b) $q$, (c) $N_0$ and (d) $\langle p_T\rangle$, extracted from the Tsallis and Hydro+Tsallis models, as a function of MCs and $dN_{ch}/d\eta$. The solid fit lines are the results of different models, indicated in each plot.}
\end{figure*}

\begin{figure*}
\centering
\includegraphics[width=0.49\textwidth]{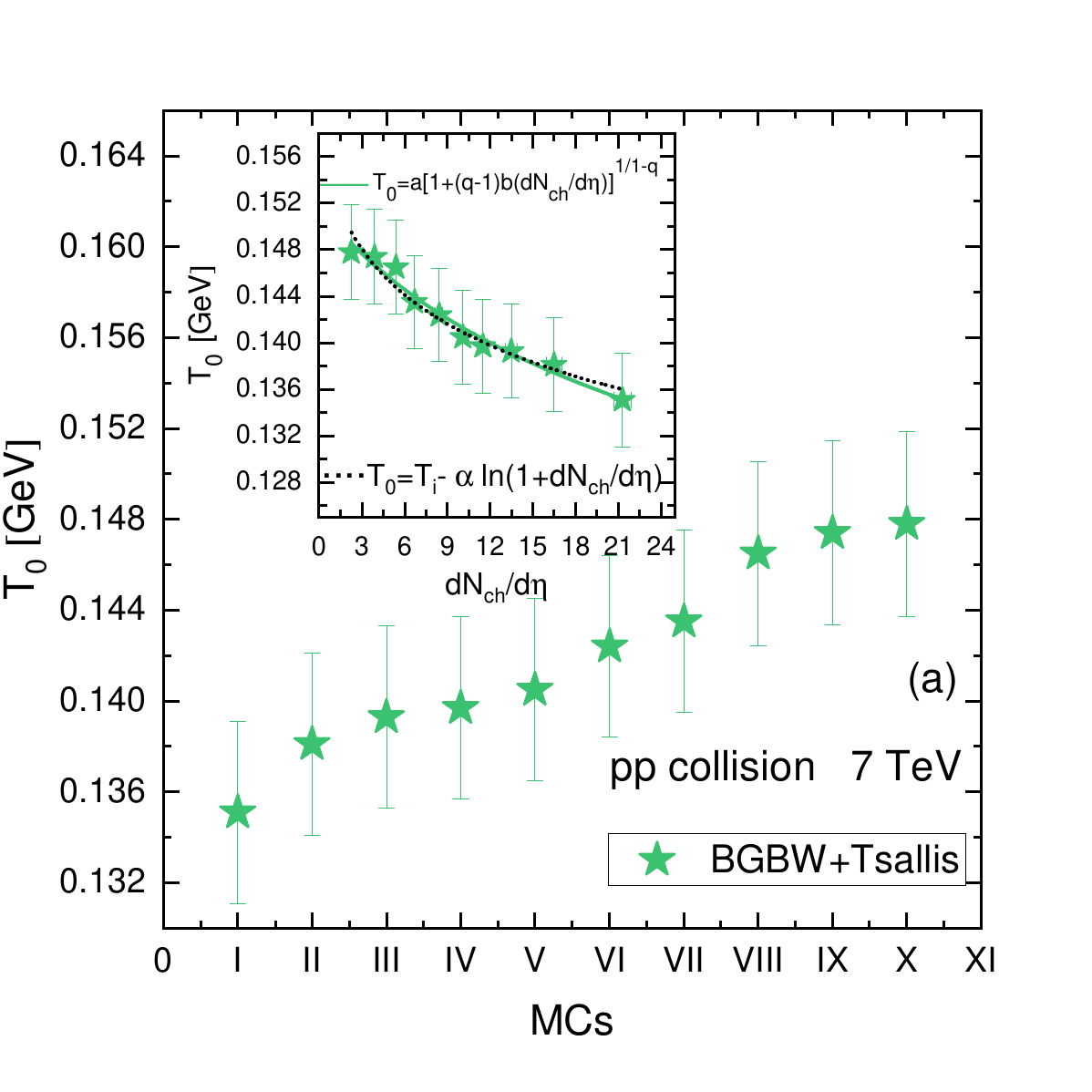}
\includegraphics[width=0.49\textwidth]{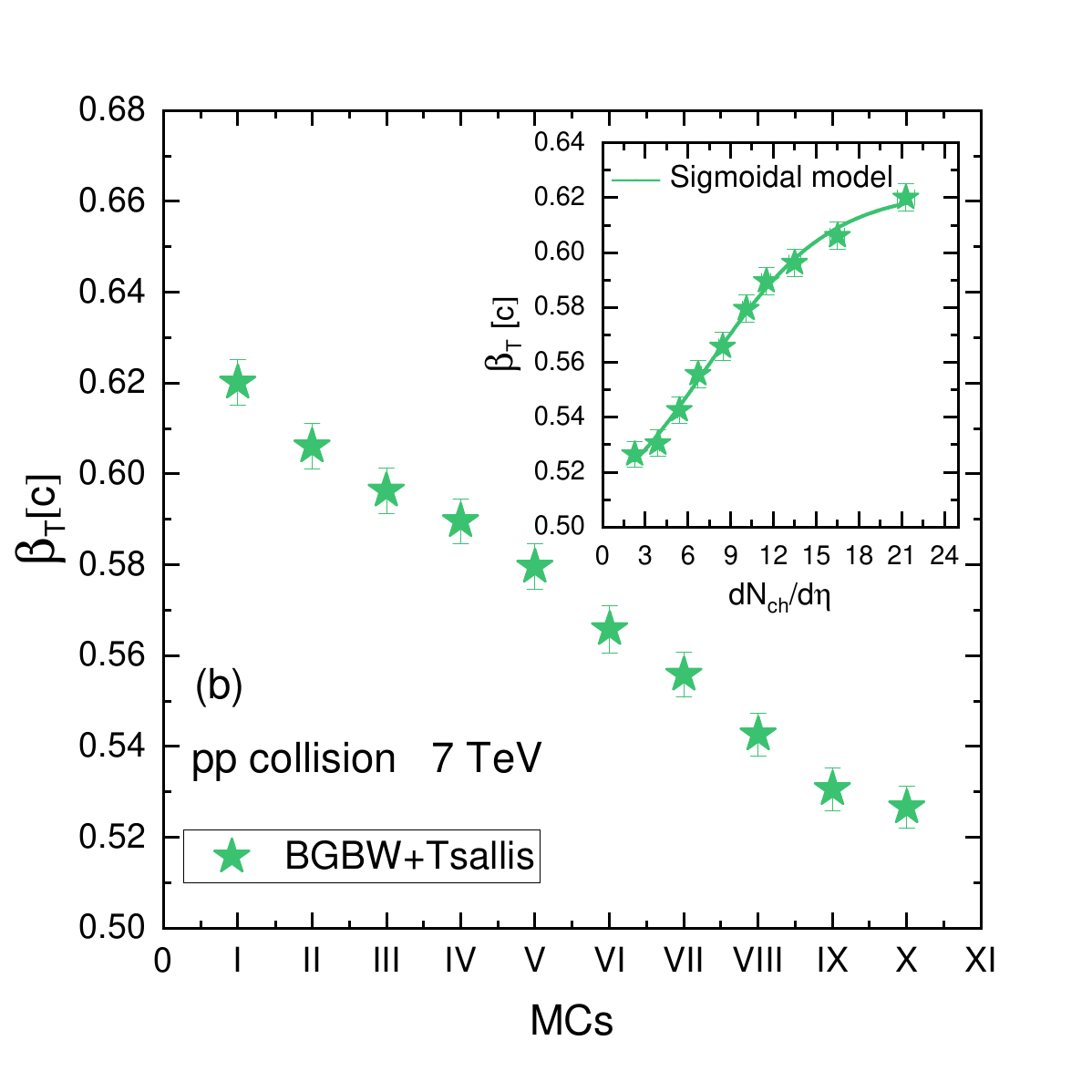}\vspace{-0.35cm}
\includegraphics[width=0.49\textwidth]{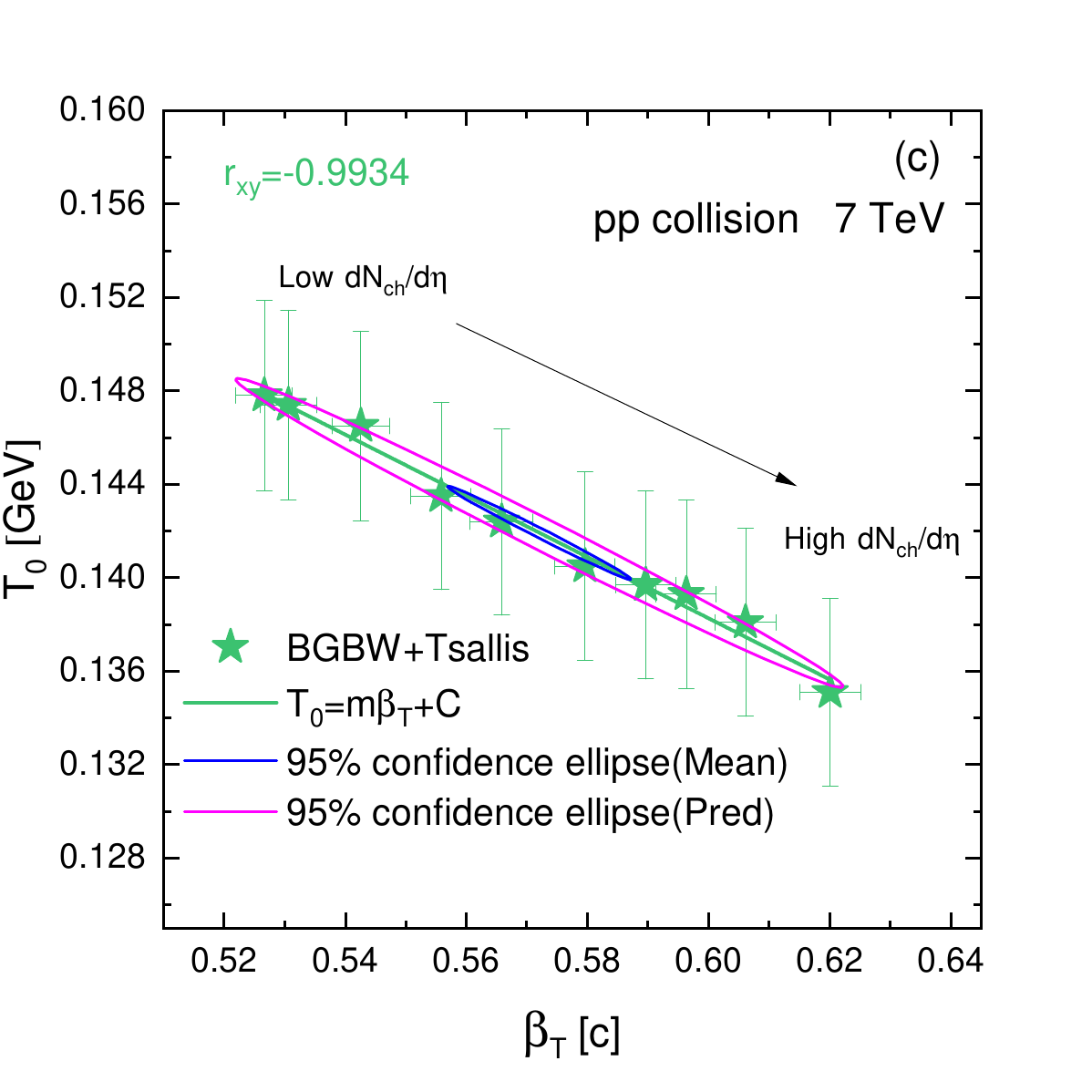}
\caption {(a) $T_0$ and (b) $\beta_T$, obtained from the Hydro+Tsallis model, as a function of MCs and $dN_{ch}/d\eta$. Plot (c) displays the correlation between $T_0$ and $\beta_T$. Different solid and dotted lines are used for the fitting results of different models, indicated in each plot.}
\end{figure*}

Fig. 3(a) represents the $T$ as a function of MCs while the inset plot of the figure represents $T$ versus $dN_{ch}/d\eta$. It is clear from the figure that with increasing(decreasing) $dN_{ch}/d\eta$(MCs), $T$ also increases, it is because higher $dN_{ch}/d\eta$ results from the higher energetic collisions where a greater amount of energy is deposited in the collision zone and hence in the produced system which leads to the greater excitation level of the system. The distinct colours and symbols are used for the two different models used. The $T$--$dN_{ch}/d\eta$ parameter space of the Tsallis model has been fitted through the logarithmic function $T=a\ln(1+b(dN_{ch}/d\eta))$, where $a=0.0396\pm 0.0011$ and $b=8.4974\pm 1.0550$ with reduced $\chi^2$ $(\chi^2/NDF)=2.93\times 10^{-5}$. The physical significance of the logarithmic model is to capture the asymptotic change, meaning that $T$ increases sharply at lower multiplicities and then slows down a little bit as multiplicity increases. This slow increase in $T$ at higher multiplicities suggests the slight saturation in $T$, which might suggest the production of a more thermalized and equilibrated system, akin to QGP. $T$ versus $dN_{ch}/d\eta$ relationship extracted from the Hydro+Tsallis model follows a linear function $T=m(dN_{ch}/d\eta)+C$ where slope $(m)=0.0029\pm 1.34\times 10^{-4}$ and intercept $(C)=0.1419\pm 0.0015$ with $\chi^2/NDF=2.21\times 10^{-5}$.  Pearson’s correlation coefficient ($r_{xy}$)  for this linear relation is 0.9913 which shows a strong positive dependence of $T$ on multiplicity density. The ellipse represents the $95\%$ confidence predicted ellipse, the ellipse is highly elongated which shows the strong linear relation of $T$ and $dN_{ch}/d\eta$. 

Fig.~3(b) displays $q$ as a function of MCs while the inset plot shows the same versus $dN_{\rm ch}/d\eta$. It can be observed that $q$ decreases (increases) with increasing $dN_{\rm ch}/d\eta$ (MCs), indicating values closer to thermal equilibrium in higher-$dN_{\rm ch}/d\eta$ events. We emphasize that the approach $q\to 1$ is suggestive but not a sufficient condition for hydrodynamics in $pp$ collisions; rather, it is consistent with trends observed in systems exhibiting collective behavior.
%Fig. 3(b) displays $q$ as a function of MCs while the inset plot is the former versus $dN_{ch}/d\eta$ plot. It can be observed that $q$ decreases(increases) with increasing $dN_{ch}/d\eta$(MCs) confirming that the system is closer to equilibrium in smaller MCs compared to higher MCs i.e., the system is more collective and thermalized in laerger $dN_{ch}/d\eta$. 
The relation in Fig. 3(b) is described by the linear relation $q=-5.43\times 10^{-4}\pm 1.12\times 10^{-5}dN_{ch}/d\eta+1.1826\pm1.29\times 10^{-4}$ with $\chi^2/NDF=4.01\times 10^{-8}$ and $q=-4.21\times 10^{-4}\pm 2.59\times 10^{-5}(dN_{ch}/d\eta)+1.1846\pm 2.96\times 10^{-4}$ with $\chi^2/NDF=2.12\times 10^{-7}$ in Tsallis and Hydro+Tsallis, respectively. The magnitude of slope for Tsallis model is greater than the Hydro+Tsallis model suggesting the rapid decrease of $q$ versus $dN_{ch}/d\eta$ in the former model compared to the latter model. We have reported $r_{xy}=-0.9985$ and $r_{xy}=-0.9951$ for Tsallis and Hydro+Tsallis models, respectively. The $r_{xy}$ values indicate the stronger negative dependence of $q$ on $dN_{ch}/d\eta$ in the former model compared to the latter model. This result is also obvious from the $95\%$ confidence ellipses where data from the Tsallis model has the ellipse with a narrower dimension than the one in the Hydro+Tsallis model. The dependence of $q$ on the $dN_{ch}/d\eta$ also obeys the non-extensive Tsallis like function $q=a[1+(q'-1)b(dN_{ch}/d\eta))]^{1/(1-q')}$ where $q'$ is used to distinguish it from the $q$ obtained through the fitting of $p_T$ spectra, though the meaning of both $q$ is same i.e., the non-extensive parameter. The values of different parameters along with $\chi^2/NDF$ extracted from the fitting in Fig. 3(b) via non-extensive Tsallis-like function are tabulated in Table 5.

Fig. 3(c) is used to represent the relation of normalization constant, also known as the multiplicity parameter, ($N_0$) and MCs while the inset plot is for $N_0$ versus $dN_{ch}/d\eta$. $N_0$ is observed to decrease(increase) with increasing MCs($dN_{ch}/d\eta$) because of the smaller energy availability for new particles production in the higher MCs. The $N_0$--$dN_{ch}/d\eta$ parameters space follows the linear function with $r_{xy}=0.9988$ and $r_{xy}=0.9997$ in Tsallis and Hydro+Tsallis, respectively, showing the strong positive dependence of the former on the latter. The $r_{xy}$ values suggest the stronger increase of $N_0$ with $dN_{ch}/d\eta$ in the Hydro+Tsallis compared to the Tsallis model. See Table 4 for the fitting parameters obtained through the linear fitting.

The dependence of $\langle p_T\rangle$ on the MCs and $dN_{ch}/d\eta$ (the inset plot) is shown in Fig. 3(d). The elevated trend of $\langle p_T\rangle$ with $dN_{ch}/d\eta$ is evident from the figure, which is because of the reason that the high multiplicity events are related to high energetic collisions where maximum energy is transferred to the system, resulting in higher $\langle p_T\rangle$. The dependence of $\langle p_T\rangle$ on $dN_{ch}/d\eta$ in the Tsallis model is described by the logarithmic model $\langle p_T\rangle=a\ln(1+b(dN_{ch}/d\eta))$ which captures the asymptotic nature of the dependence, where $\langle p_T\rangle$ increases rapidly at low $dN_{ch}/d\eta$ and then started to saturate a little bit at higher $dN_{ch}/d\eta$ pointing to the possible transition towards a thermalized and hydrodynamic system more like a QCD matter. The free parameters extracted from the logarithmic model are given in Table 3. The dependence of $\langle p_T\rangle$, extracted from the Hydro+Tsallis model, on $dN_{ch}/d\eta$ follows the linear function with $r_{xy}=0.9923$, showing the strong positive dependence of the former on the latter. The $95\%$ confidence predicted ellipse is also drawn, whose elongated dimension also represents the strong relation of $\langle p_T\rangle$ with $dN_{ch}/d\eta$. The values of intercept and slope for linear fitting are tabulated in Table 4.  

Fig. 4(a) represents the kinetic freeze-out temperature as a function of MCs and $dN_{ch}/d\eta$ (the inset plot). $T_0$ declines as $dN_{ch}/d\eta$ increases suggesting the formation of short-lived fireball in the low $dN_{ch}/d\eta$. The relation is described by the logarithmic decay model $T_0=(T_0)_i-\alpha \ln(1+dN_{ch}/d\eta)$ where $(T_0)_i$ is the temperature at lowest possible $dN_{ch}/d\eta$ and $\alpha$ represents rate of decay (decrease) in $T_0$ with increasing $dN_{ch}/d\eta$. The logarithmic decay model suggests that at higher multiplicities ($dN_{ch}/d\eta$), most of the energy is incorporated in the production of the particles, leaving behind a smaller portion of energy to contribute to the system's temperature. On the other hand, at low $dN_{ch}/d\eta$, a smaller number of particles are produced and hence a comparatively larger portion of the energy is available for the system excitation. The fitted line arising from the logarithmic decay model is denoted by the dotted line in the plot, while the resulting fit parameters are $({T_0})_i=0.1578\pm0.0013$  GeV, $\alpha=0.0070\pm5.56\times10^{-4}$ with $\chi^2/NDF=6.85\times10^{-6}$. The relation in Fig. 4(a) is also described by the non-extensive Tsallis-like function, indicated by the solid curve. 

Fig. 4(b) displays transverse flow velocity $\beta_T$ as a function of MCs and $dN_{ch}/d\eta$ (the inset plot), the former increases with a decrease(increase) in the MCs($dN_{ch}/d\eta$) as maximum particles are produced in low MCs which leads to greater pressure gradient in the system and results in the higher $\beta_T$. The relation in Fig. 4(b) obeys the sigmoidal model $y=(y_0-y_{min})/(1+\exp{((x-M)/\Delta)}+y_{min}$, where $y_0$ is the lowest value of the quantity on the y-axis, here $\beta_T$, corresponding to the zero value of the parameter on the x-axis, here $dN_{ch}/d\eta$, $y_{min}$ may denote the least maximum value of $y$ corresponding to the maximum value on the x-axis, $x$ denotes the parameter on the x-axis, $M$ represents the point on the x-axis where the drop or increase in the parameter on the y-axis is more significant and $\Delta$ measures how sharply the transition occurs. The sigmoidal model is considered to be the best-fitting tool for the S-shaped data, where $\beta_T$ initially increases slowly at low $dN_{ch}/d\eta$, then increases sharply in some specific intermediate $dN_{ch}/d\eta$ interval and then slows down or even saturates at higher $dN_{ch}/d\eta$. The slower increase in $\beta_T$ at higher $dN_{ch}/d\eta$ after a sharp increase in the intermediate $dN_{ch}/d\eta$ may indicate the transition from a non-collective medium to a collective hydrodynamic medium. The fit parameters extracted along with $\chi^2/NDF$ values are tabulated in Table 6.

The correlation between $T_0$ and $\beta_T$ is plotted in Fig. 4(c). This correlation follows the linear relation, where $r_{xy}=-0.9934$ represents a strong negative correlation between $T_0$ and $\beta_T$. This negative correlation may be explained as, at higher $dN_{ch}/d\eta$, a large portion of the collision energy is utilized in the formation of a particle-rich and collective system, having a larger pressure gradient resulting in the higher $\beta_T$, leaving behind smaller energy for the system excitation. The $95\%$ confidence predicted and mean ellipses have also been drawn confirming the stronger linear correlation.    

\begin{figure*}
\centering
\includegraphics[width=0.45\textwidth]{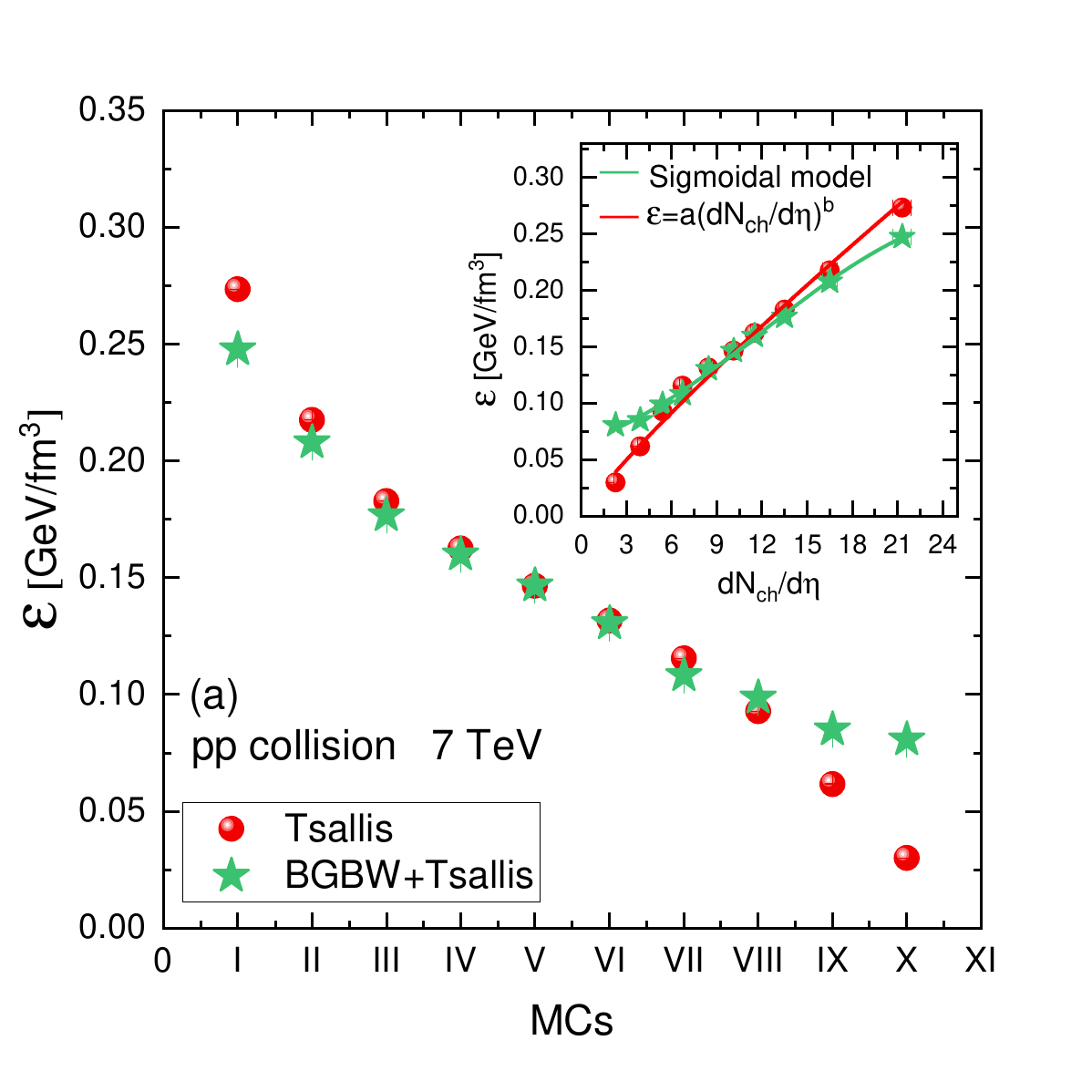}
\includegraphics[width=0.45\textwidth]{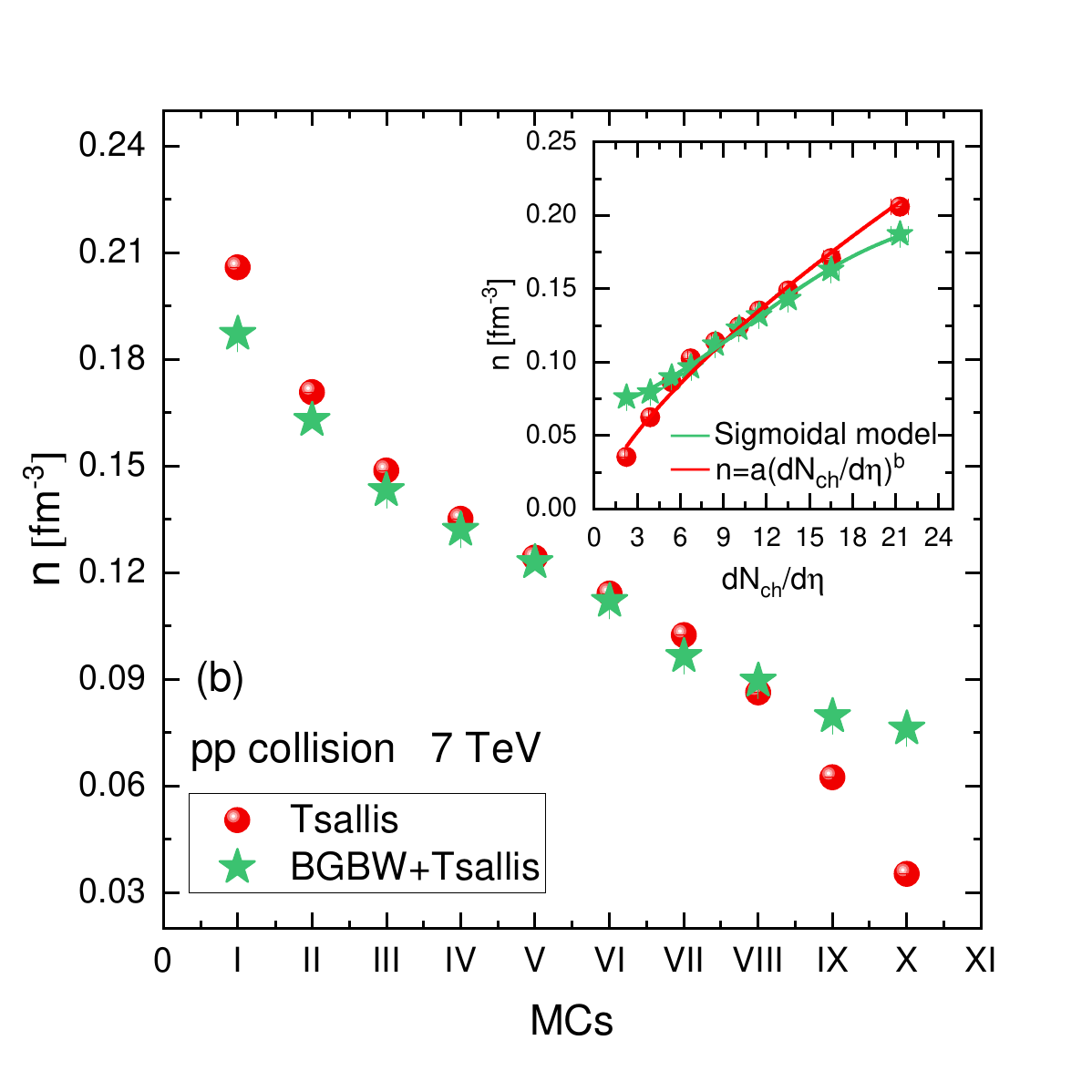}\vspace{-0.35cm}
\includegraphics[width=0.45\textwidth]{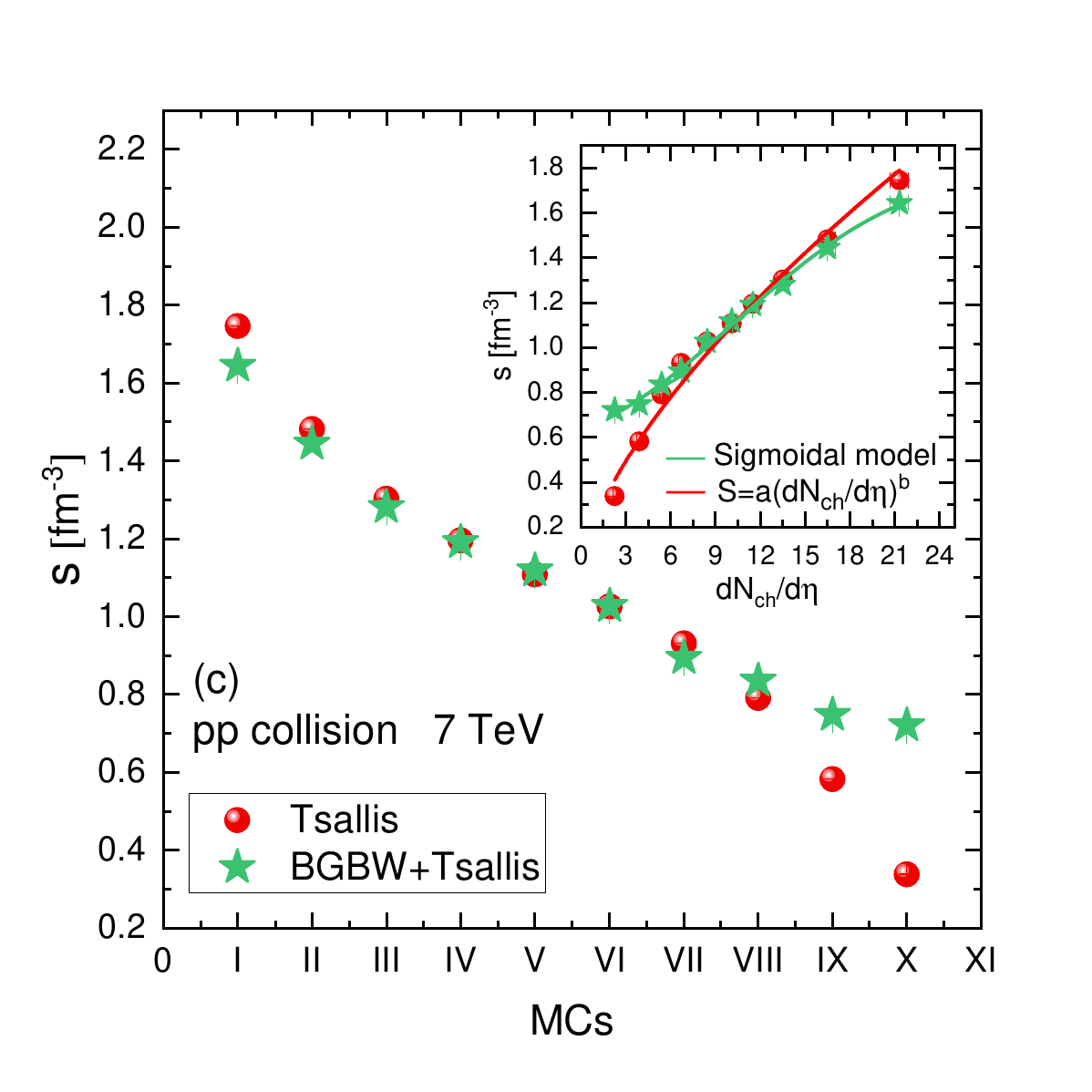}
\includegraphics[width=0.45\textwidth]{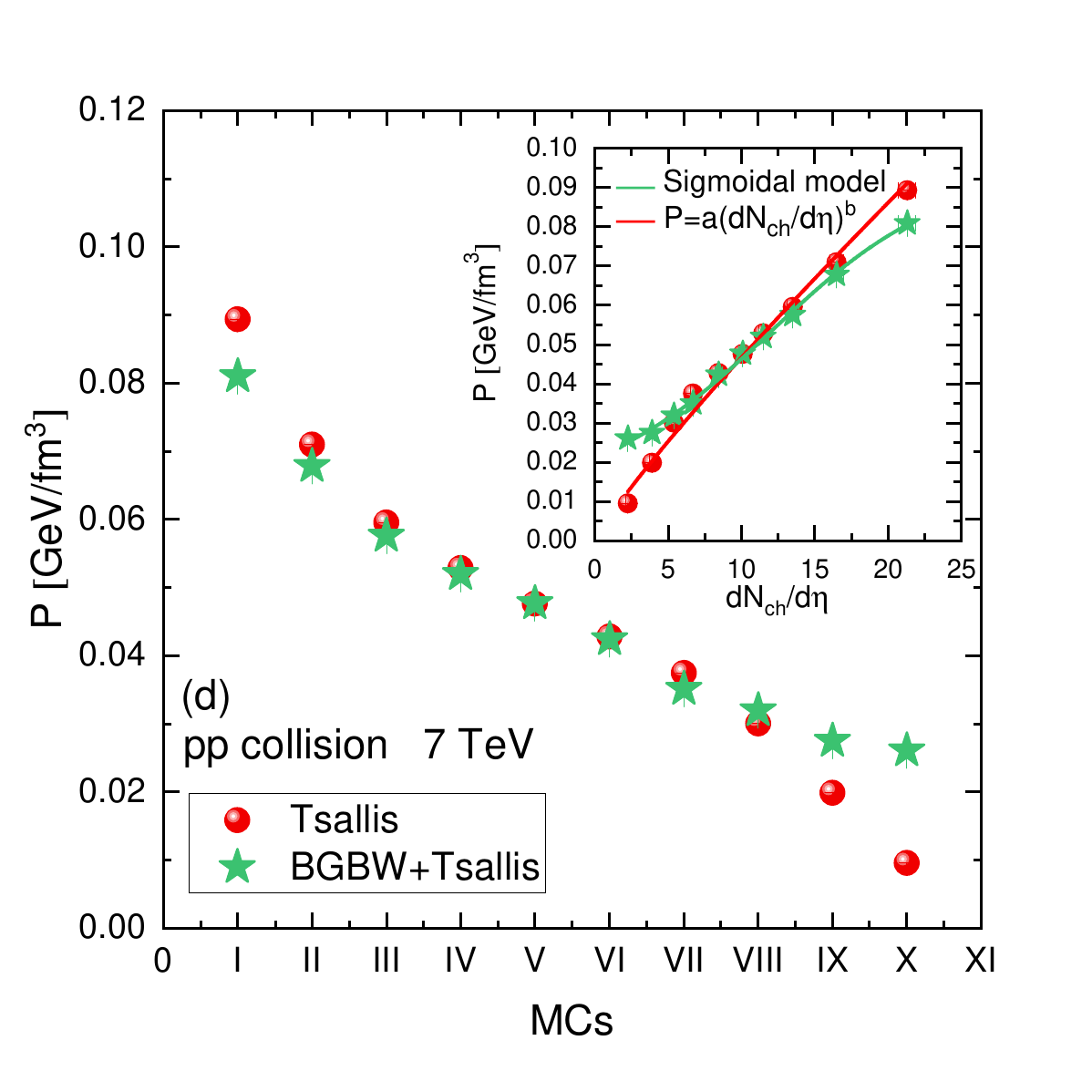}\vspace{-0.35cm}
\includegraphics[width=0.45\textwidth]{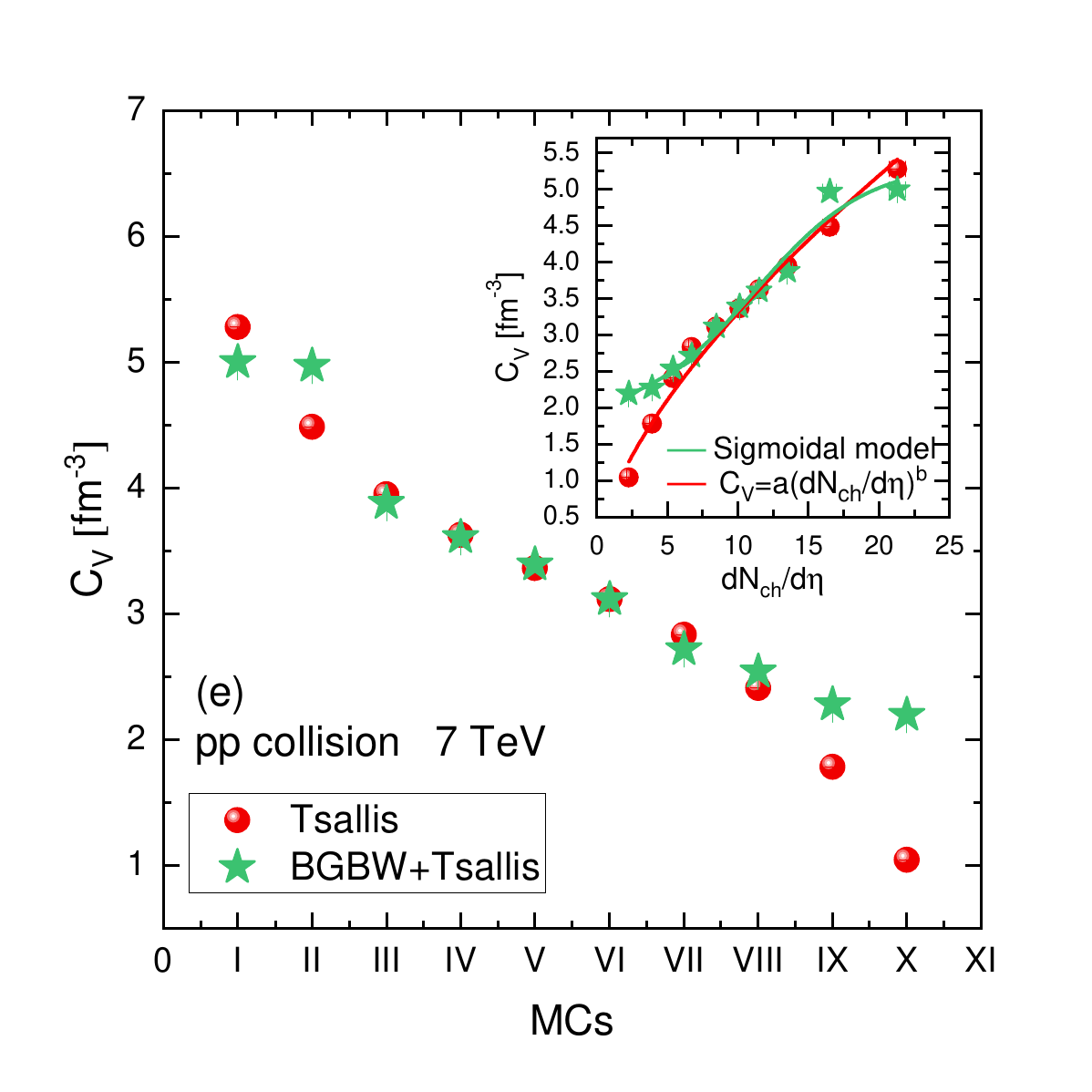}
\includegraphics[width=0.45\textwidth]{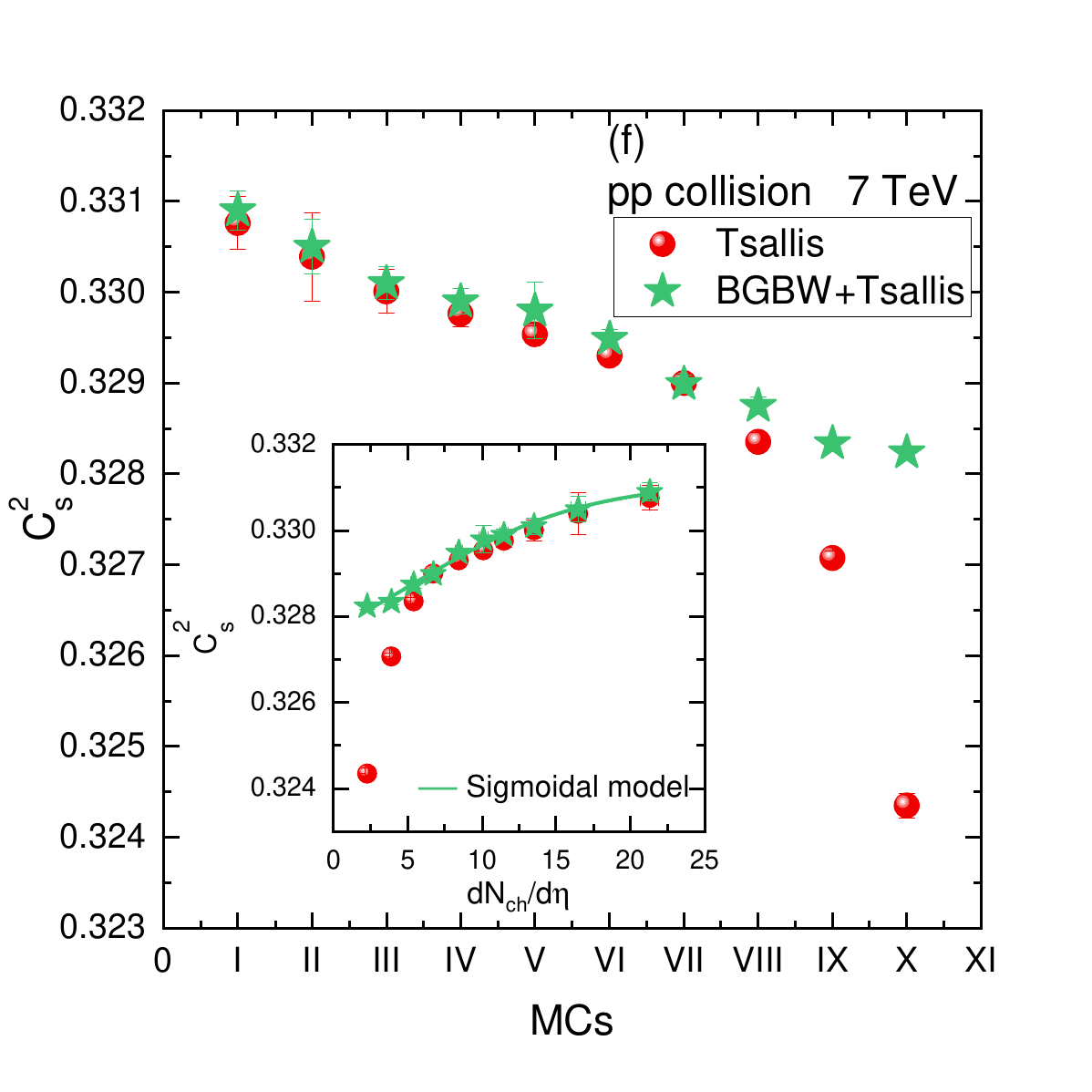}\vspace{-0.35cm}
\caption {(a) $\varepsilon$, (b) $n$, (c) $s$, (d) $p$, (e) $C_V$, and (f) ${c_s}^2$, calculated from the Tsallis and Hydro+Tsallis models, as a function of MCs and $dN_{ch}/d\eta$. The solid and dotted lines are used for the fit results of different models.}
\end{figure*}

\begin{figure*}
\centering
\includegraphics[width=0.45\textwidth]{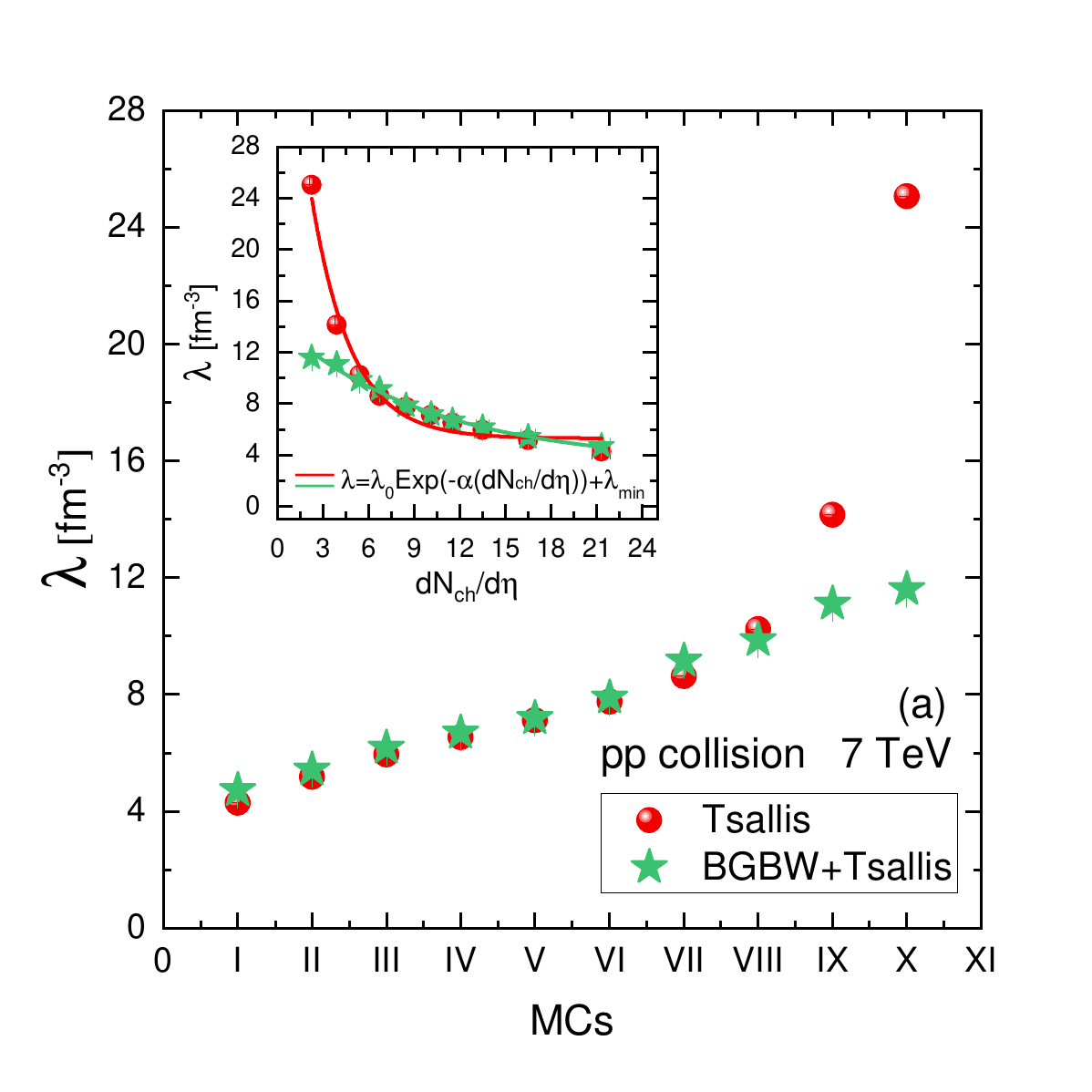}
\includegraphics[width=0.45\textwidth]{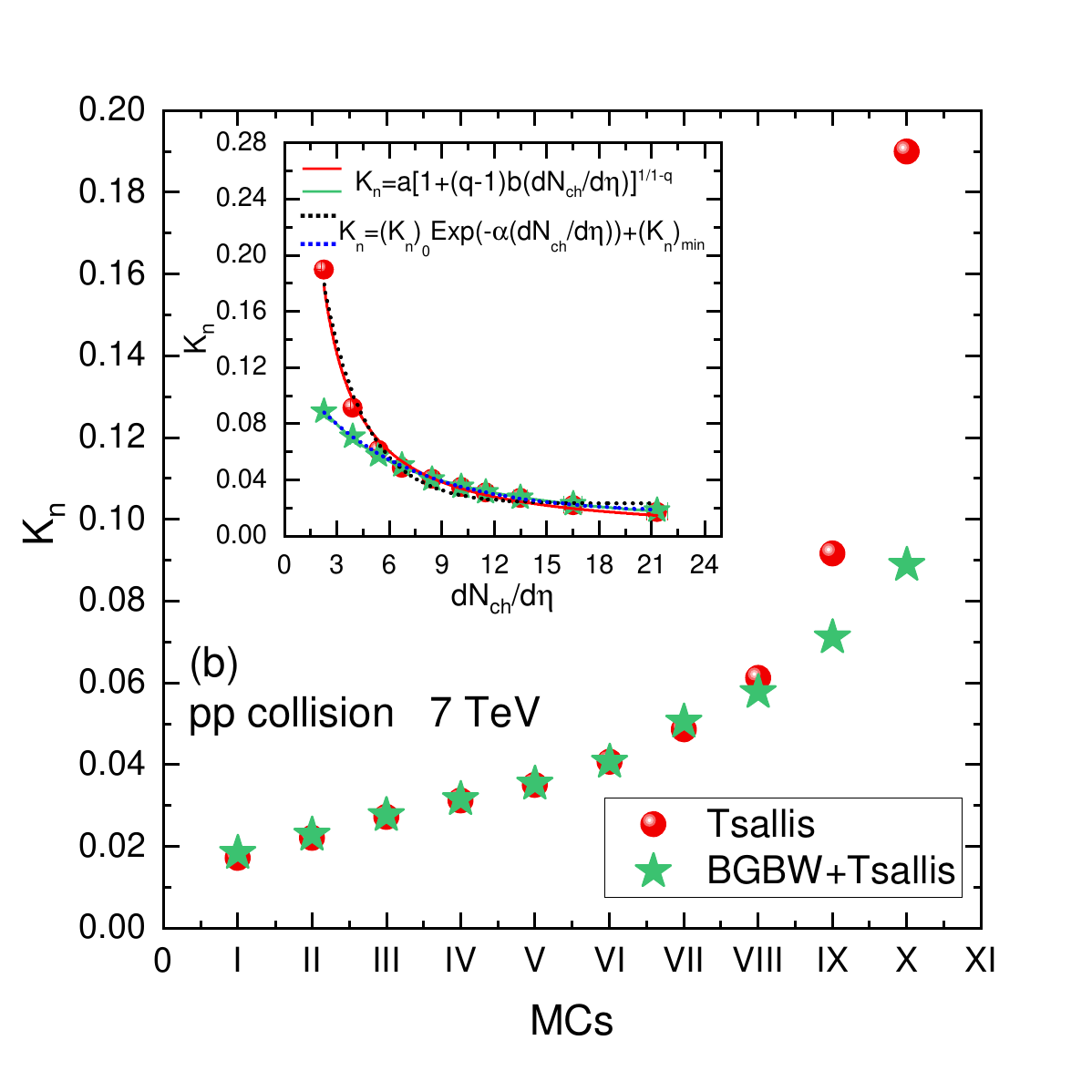}\vspace{-0.35cm}
\includegraphics[width=0.45\textwidth]{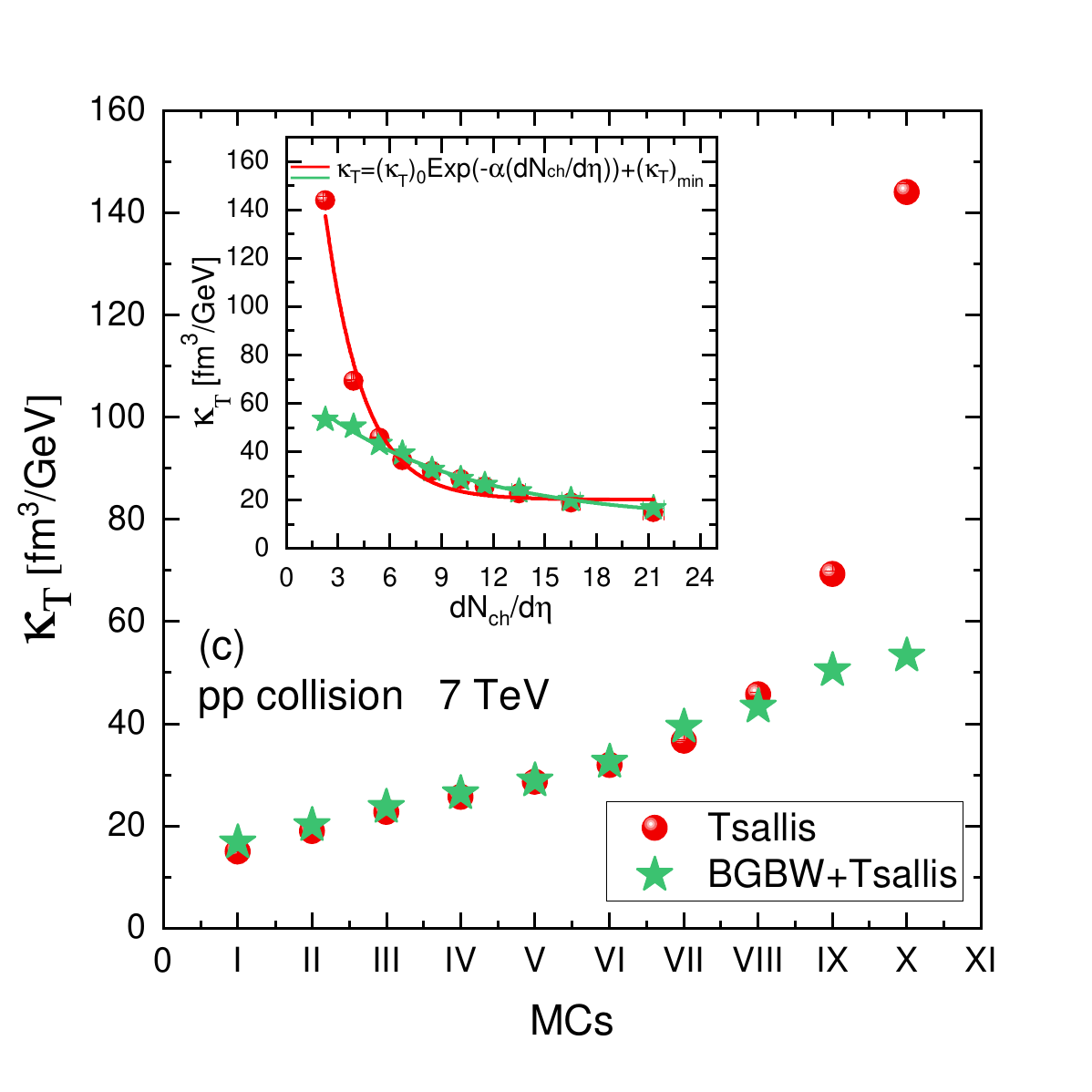}
\includegraphics[width=0.45\textwidth]{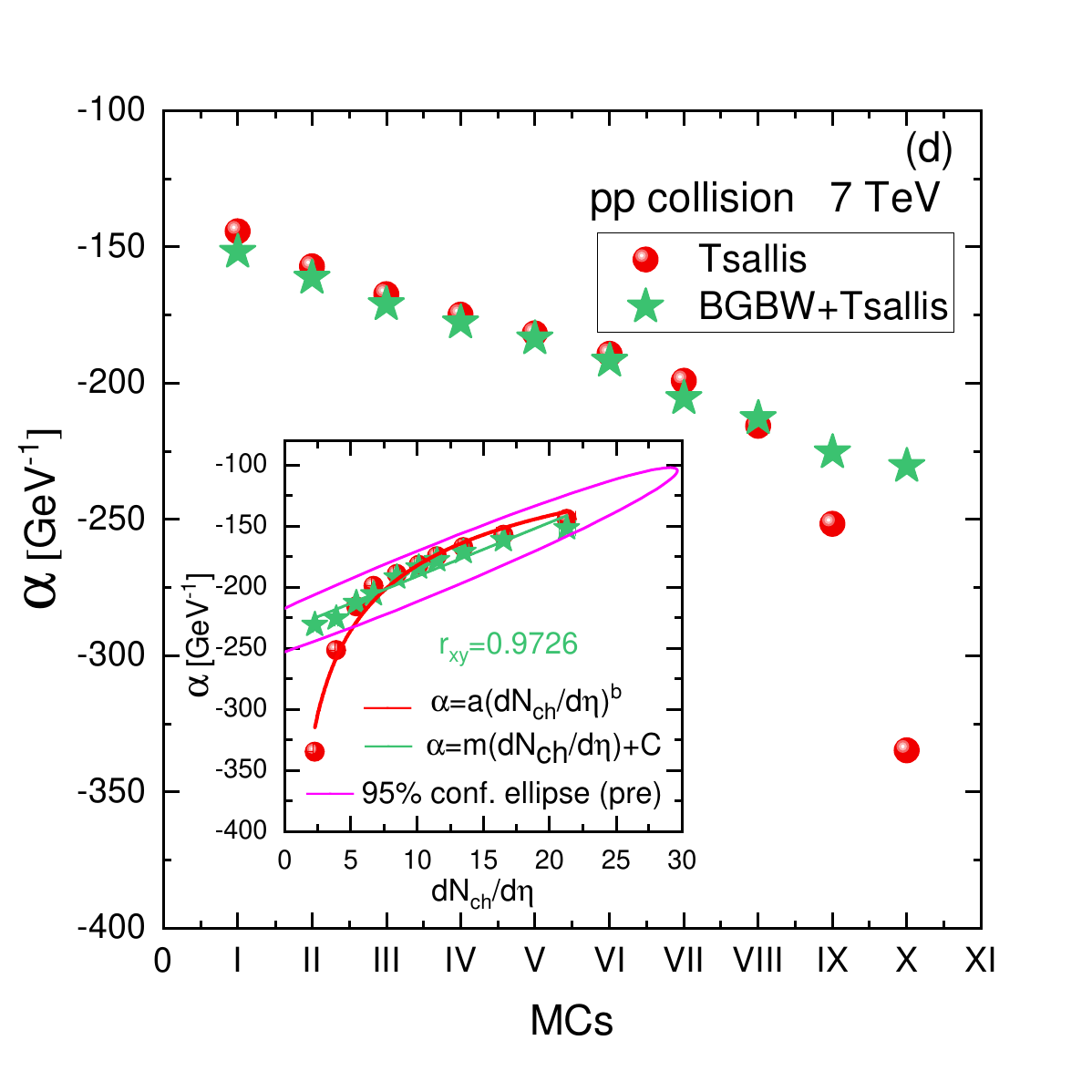}\vspace{-0.35cm}
\caption {(a) $\lambda$, (b) $K_n$, (c) $\kappa_T$, and (d) $\alpha$, calculated from Tsallis and Hydro+Tsallis models, versus MCs and $dN_{ch}/d\eta$. The solid and dotted lines are used for the fit results of different models.}  
\end{figure*}

Fig. 5(a) represents $\varepsilon$ versus MCs, while the inset is the plot between the former and $dN_{ch}/d\eta$. One can see that with increasing(decreasing) $dN_{ch}/d\eta$(MCs), $\varepsilon$ increases because of the intense collisions in the lower MCs. Fig. 5(a) follows the sigmoidal model for Hydro+Tsallis model and power law $\varepsilon=a(dN_{ch}/d\eta)^b$ for pure Tsallis. The values of different free parameters extracted from the sigmoidal model and power law fitting, along with $\chi^2/NDF$ are given in Tables 6 and 7, respectively.

Fig. 5(b) shows an elevated trend of the particle number density $n$ with decreasing(increasing) MCs($dN_{ch}/d\eta$). The maximum energy availability for the charged particle production in the lower MCs(higher $dN_{ch}/d\eta$) may be the possible reason for the elevated values of $n$ in the lower MCs. Fig. 5(b) follows the sigmoidal model for Hydro+Tsallis model and power law for pure Tsallis. 

The varying trend of entropy density $s$ with MCs and $dN_{ch}/d\eta$(inset plot) is shown in Fig. 5(c), where $s$ is observed to be increasing (decreasing) with decreasing MCs($dN_{ch}/d\eta$) as low MCs are the results of the high energetic collision which leads to the production of a comparatively thermodynamically active or disordered system. $s$--$dN_{ch}/d\eta$ parameter space is described by the sigmoidal model for Hydro+Tsallis and power law for Tsallis model.

Fig. 5(d) displays the increasing trend of pressure $P$ with decreasing MC, while the inset plot shows the growing trend of $P$ with increasing $dN_{ch}/d\eta$. The low MCs are associated with harsh collisions, resulting in higher pressure in the collision zone and the produced system. Again, the variations in $P$ concerning changing $dN_{ch}/d\eta$ are described by the sigmoidal model for Hydro+Tsallis and the power law for the pure Tsallis model.

Fig. 5(e) presents the specific heat capacity at constant volume $C_V$ as a function of the MCs and $dN_{ch}/d\eta$ (the inset plot). $C_V$ increases with decreasing MCs as low MCs have higher $dN_{ch}/d\eta$, which results in the increased interaction rates among the constituents of the produced system and in the production of a system with higher energy density having the larger capacity to store and transfer heat. The dependency of $C_V$ over $dN_{ch}/d\eta$ in the plot obeys the sigmoidal function and power law function for Hydro+Tsallis and Tsallis models, respectively.

The squared speed of sound ${c_s}^2$ as a function of MCs and $dN_{ch}/d\eta$ (the inset plot) is shown in Fig. 5(f). ${c_s}^2$ is related to the equation of state (EOS), especially to the one connecting the energy density and pressure. The larger value of ${c_s}^2$ suggests the stiffer equation of state, i.e., the production of the less compressible medium, which supports the propagation of sound waves. In Fig. 5(f), we reported the increasing trend of ${c_s}^2$ with decreasing(increasing) MCs($dN_{ch}/d\eta$), possibly confirming the production of a stiffer and non-compressible system more like the QGP droplet in the lower MCs(higher multiplicity). The values of ${c_s}^2$ are very close to $1/3$ as reported in the previous studies for the pion gas \cite{deb2021study, sahu2020multiplicity, sahoo2024multiplicity, deb2020deciphering}. The relation in Fig. 5(f) also follows the sigmoidal model for the Hydro+Tsallis model.

Fig. 6(a) displays the mean free path $\lambda$ versus MCs and $dN_{ch}/d\eta$ (the inset plot). $\lambda$ is the mean distance a particle covers between two successive collisions. With increasing $dN_{ch}/d\eta$ or decreasing MCs the system's energy and particle densities increase, resulting in the smaller distances among the particles of the produced system and eventually in the reduced $\lambda$. $\lambda$ versus $dN_{ch}/d\eta$ relation obeys the exponential decay model (function) $\lambda=\lambda_0\exp{(-\alpha(dN_{ch}/d\eta))}+\lambda_{min}$. Where $\lambda_0$ denotes $\lambda$'s greater value at almost zero $dN_{ch}/d\eta$, $\lambda_{min}$ is the minimum value of $\lambda$ at the greater $dN_{ch}/d\eta$ in the plot while $\alpha$ represents the rate of decay (decrease) in $\lambda$ concerning $dN_{ch}/d\eta$. Here, $\alpha$ should not be mixed with the expansion coefficient. The different extracted parameters and the reduced $\chi^2$ values are given in Table 8. 

The Knudsen number $K_n <$ 1 represents the system with a higher particle density, where the mean free path of the particles is smaller than the system’s size, where the collisions among the constituents of the system are frequent, bringing the system to the quick local thermal equilibrium and to behave more like a hydrodynamical system akin to QGP. $K_n >$ 1 indicates a system where the mean free path is larger than the system’s size, resulting in a more dilute system where the collective properties are absent, akin to a hadron gas. Fig. 6(b) presents $K_n$ as a function of MCs and $dN_{ch}/d\eta$ (inset plot). One can observe that with decreasing(increasing) the MCs($dN_{ch}/d\eta$), $K_n$ also decreases, which may point to the production of a more thermalized, locally equilibrated and collective medium more like QCD matter. The relation in Fig. 6(b) follows the exponential decay model denoted by the solid lines; the free parameters and $\chi^2/NDF$ values of this fit are tabulated in Table 8. The relation can also be described using a non-extensive Tsallis-like model denoted by the dotted line, see Table 5.

It must be noted that the $K_n <$ 1 condition, which is found in all multiplicity classes, including low multiplicity pp events, is not sufficient to obtain a fully equilibrated or long-lived hydrodynamic medium that is analogous to that of nucleus-nucleus collisions. Rather, since the calculation of $K_n$ at the kinetic freeze-out point is performed using thermodynamic quantities calculated with Tsallis statistics, $K_n <$ 1 implies that the mean free path at freeze-out is smaller than the characteristic system size, and that the dynamics at freeze-out are interaction-dominated and locally collective. This interaction is in agreement with the notion of hydrodynamization at non-complete thermal equilibrium that has been extensively studied in the case of small systems at LHC energies. Thus, the measured Knudsen number is evidence of effective collectivity and not true macroscopic fluid behavior.

Fig. 6(c) displays the decreasing trend of isothermal compressibility $\kappa_T$ with decreasing(increasing) MCs($dN_{ch}/d\eta$). The decreasing $\kappa_T$ with increasing $dN_{ch}/d\eta$ might suggest the transition to a locally thermalized, equilibrated, and hydrodynamic (collective) medium, more like a liquid, from a dilute and a non-hydrodynamic medium, more like a gas, in low $dN_{ch}/d\eta$. An exponential decay model characterises the variation of $\kappa_T$ with $dN_{ch}/d\eta$. See Table 8 for the free parameters and $\chi^2/NDF$ extracted from the fit of the exponential decay model.  

Fig. 6(d) presents expansion coefficient $\alpha$ versus MCs and $dN_{ch}/d\eta$. Where $\alpha$ measures how much the produced system's volume changes with temperature changes at constant pressure. After the creation of QGP, it expands, and its temperature falls till hadronization. Suppose one looks in the reverse direction from hadronization till the onset of QGP, then the system contracts as its temperature increases. In that case, the negative sign of $\alpha$ in our analysis represents the reduction of volume with elevation in temperature. The elevated trend of $\alpha$ with increasing $dN_{ch}/d\eta$ implies that as more particles are produced in the collision, the system's ability to expand with temperature also becomes significant. The relationship shown in Fig. 6(d) derived from the Tsallis model follows a power-law behaviour $\alpha=a(dN_{ch}/d\eta)^b$ (see Table 7 for fit parameters), while a linear pattern characterizes the relationship obtained from the Hydro+Tsallis model, see Table 4. For the linear relation, $r_{xy}=0.9726$, which shows a strong positive linear dependence of $\alpha$ on $dN_{ch}/d\eta$, this strong linear dependence is also obvious from the narrow dimension of the $95\%$ confidence ellipse.

In the present work, all quantities are obtained within a Tsallis description of pion transverse momentum spectra. The large set of derived quantities ($\varepsilon$, $n$, $s$, $P$, $C_V$, ${c_s}^2$, $\alpha$, $\lambda$, $K_n$, and $\kappa_T$) might risk overwhelming the readers; therefore, we focus on a single benchmark observable, ${c_s}^2$. For charged pions, ${c_s}^2$ does not show any strong dependence on the multiplicity of the event and is close to the ideal gas limit, with values of the order of $0.33$. This behavior is in agreement with previous studies of Tsallis in pp collisions \cite{deb2021study, sahu2020multiplicity, sahoo2024multiplicity, deb2020deciphering}, which also report that ${c_s}^2$ approaches 1/3 in high multiplicity.

At the same time, the absolute value of ${c_s}^2$ is model-dependent in the case of small systems. Hadron resonance gas (HRG) calculations give smaller values (${c_s}^2\simeq0.24$) \cite{sarwar2017estimate}, and lattice calculations of equations of state exhibit a strong decrease of ${c_s}^2$ in the vicinity of the crossover region. The larger values of ${c_s}^2$ that we get here are simply due to the effective stiffness of the freeze-out pion distribution in the Tsallis scheme and not a universal fact about equilibrated QCD matter.

We therefore make the distinction between robust and model-dependent aspects of our results. The dependence of ${c_s}^2$ on multiplicity is determined to be strong in the sense that the dependence has a direct correlation to the measured pion spectra. In contrast, the absolute values of ${c_s}^2$ and others of derived quantities, especially those with additional assumptions such as system size or interaction cross sections, must be considered as model-sensitive quantities and are interpreted in a qualitative way only regarding pp collisions.

\begin{figure*}
\centering
\includegraphics[width=0.45\textwidth]{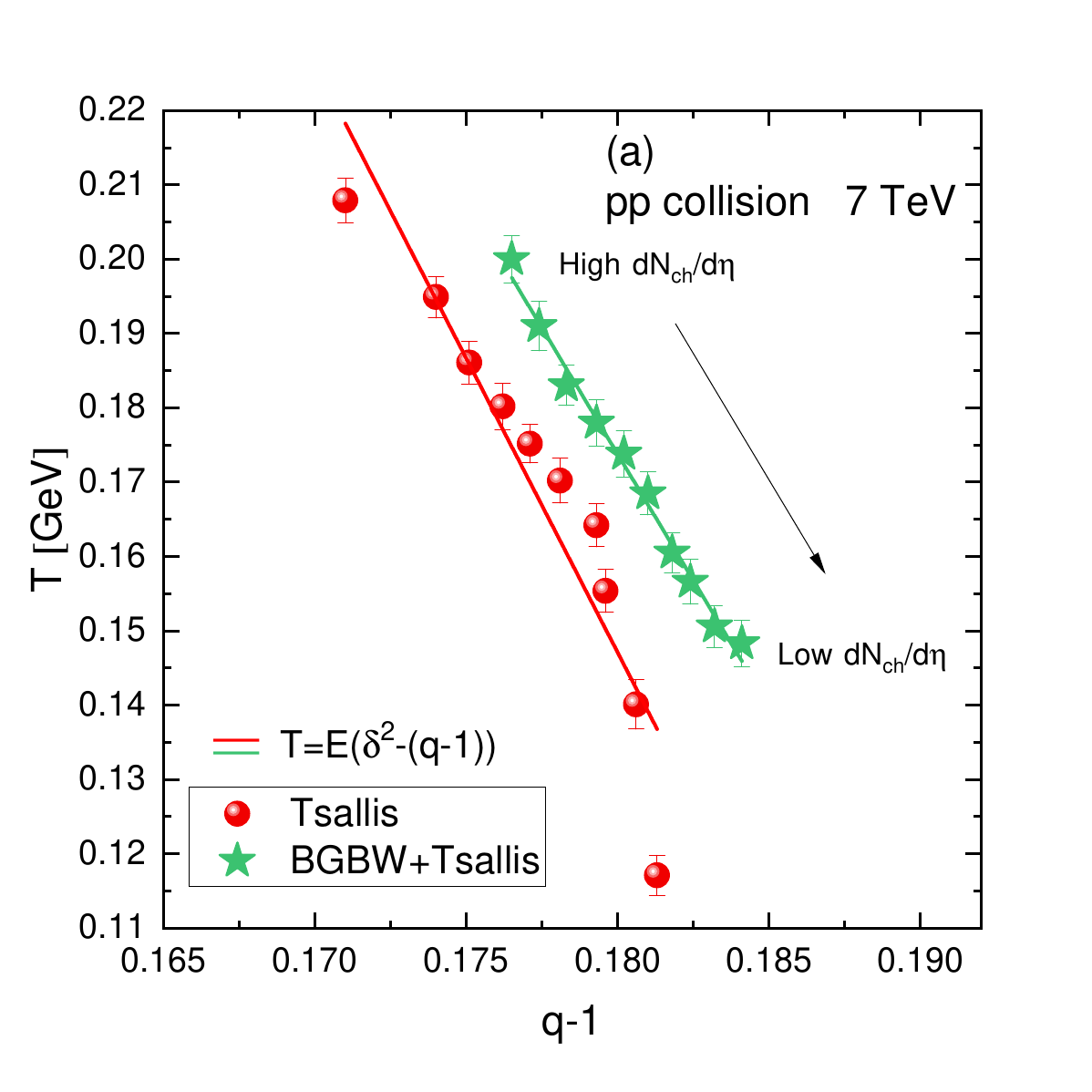}
\includegraphics[width=0.45\textwidth]{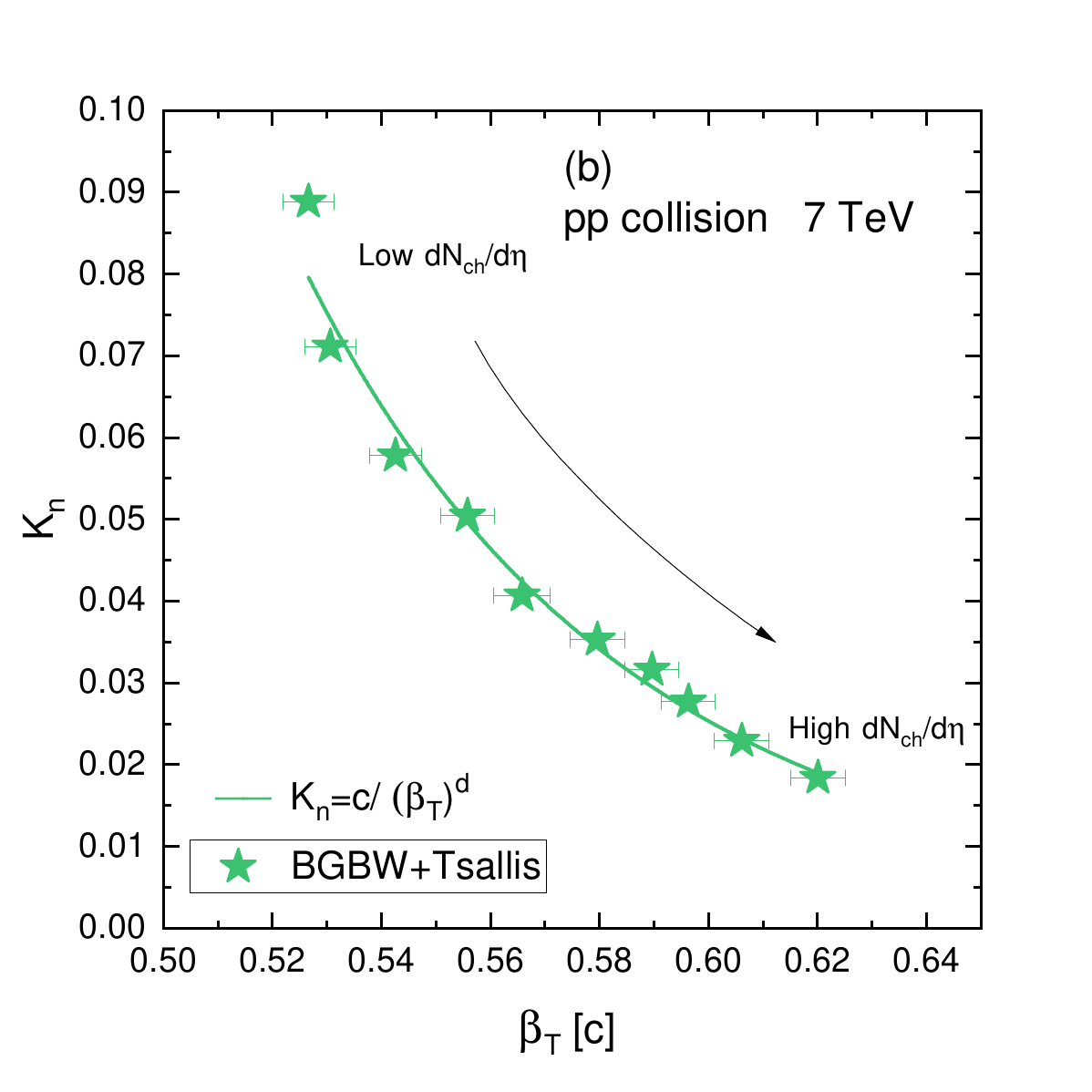}\vspace{-0.35cm}
\caption {The correlation between (a) $T$ and $q-1$ and (b) $K_n$ and $\beta_T$. The solid lines are the fit results of different functions indicated in each plot.}
\end{figure*}

Fig. 7(a) shows the correlation between $T$ and $q-1$, indicating that the former decreases with increasing the latter. This relationship shows that the system temperature increases as $q-1$ approaches 0 or $q$ approaches unity. In other words, initially, when the system is hot enough, due to the smaller mean free path, the constituents of the system frequently collide, possess a collective nature, and have a well-established local thermal equilibrium. The $T-(q-1)$ parameter space is described by the relation: $T=E(\delta^2-(q-1))$ \cite{su2021non}, where $E$ and $\delta$ are the free parameters indicating the associated energy and strength of the relationship in $T-(q-1)$ parameter space, respectively. In our case, we obtain $E=6.7919\pm 0.2429$ and $\delta=0.4534\pm 0.0010$ with $\chi^2/NDF=12.1232$ in the Tsallis model, while for Hydro+Tsallis model, $E=7.9160\pm 1.0559$ and $\delta=0.4456\pm 0.0032$ with $\chi^2/NDF=0.3664$.    

Fig. 7(b) shows the negative correlation between $K_n$ and $\beta_T$. With the increase in $\beta_T$, $K_n$ decreases, which may suggest that for the multiplicity classes with heightened values of $\beta_T$, the system exhibits a collective nature, as evident from the low $K_n$ values. The correlation between $K_n$ and $\beta_T$ obeys the power law equation $K_n=c/{\beta_T}^d$ where c and d are the free parameters having values of $c=2.86E-4\pm 5.18E-5$ and $d=8.7742\pm 0.3183$, respectively, with $\chi^2/NDF=0.0103$. 
\\
\subsection{Further Discussion}
Collective-like behavior at 7 TeV pp collisions has been explored by previous studies in terms of Tsallis-based statistical descriptions and blast-wave-inspired models. Specifically, pion spectra collected by ALICE were fitted with independent Tsallis and Boltzmann-Gibbs Blast-Wave (BGBW) fits, and the multiplicity dependence of the obtained parameters was presented in Ref. \cite{khuntia2019radial}. These analyses were then extended to derive thermodynamic quantities like the average transverse momentum, speed of sound, specific heat, and isothermal compressibility by assuming that the Tsallis distribution is the freeze-out phase-space distribution \cite{deb2021study, sahu2020multiplicity}. Even more recently, observables associated with transport, such as the Knudsen number, were investigated in the heavy-flavor sector based on model-dependent approximations of system size and density \cite{sahoo2024multiplicity}.

These previous studies are generalized to the current work using a two-component analysis of the complete transverse-momentum spectrum. In contrast to these three references \cite{deb2021study, sahu2020multiplicity, khuntia2019radial}, which use either the Tsallis or blast-wave model individually and frequently within a limited interval of $p_T$, we simultaneously fit the whole measured $p_T$ range with a hybrid function composed of a flowing thermal (BGBW) term in the soft sector and a Tsallis-Pareto distribution in the hard tail. The method allows the collective-flow parameters and non-extensive parameters to be self-consistently obtained as a result of a single fit, without an artificial division of soft and hard momentum space.

The main implication of such a hybrid framework is that it allows us to separate bulk collective effects and non-equilibrium contributions of hard processes. The non-extensive parameter $q$ in the previous single-component Tsallis analysis is an effective parameter to consider both soft and hard physics at the same time \cite{deb2021study, sahu2020multiplicity,khuntia2019radial}. Conversely, in the current scheme, the Tsallis component is largely used to describe the high-$p_T$ sector, and it gives a cleaner interpretation of non-extensivity and its dependence on the charged-particle multiplicity. This enables a correlated study of the bulk freeze-out parameters ($T_0$, $\beta_T$) and non-extensive parameters ($T$, $q$), which shows trends not obtainable in single-component descriptions.

Moreover, this work significantly extends the scope of thermodynamic and transport observables that have been extracted in a single framework, which is internally consistent. Although Refs. \cite{deb2021study, sahu2020multiplicity} calculated thermodynamic quantities by taking a Tsallis distribution as the only freeze-out distribution; we calculate these observables by taking the Tsallis component constrained by the system size (effective source size at kinetic freeze-out) and low-$p_T$ yields obtained as a result of the blast-wave fit. This results in better estimates of particle density, entropy, and mean free path at kinetic freeze-out.

Physically, the dependence on multiplicity of the parameters extracted, i.e., growing transverse flow velocity, decreasing non-extensivity parameter, growing particle density, and decreasing Knudsen number, may be understandable in terms of the development of more active collective dynamics in high-multiplicity pp collisions. These trends may be associated with situations when there are increased multiparton interactions, color reconnection, large initial parton densities, or final-state rescattering, all of which can produce the appearance of collective-like behavior without necessarily having to form a long-lived deconfined medium \cite{deb2021study, sahu2020multiplicity, sahoo2024multiplicity, khuntia2019radial}. In this context, the current hybrid analysis is a unique and quantitatively constrained description of soft and hard physics, which provides a new understanding of how collective behavior in small collision systems may form.

It is important to make it clear that the present analysis does not aim at the unambiguous formation of deconfined QGP in small systems of collision. In particular, the rise in average transverse momentum and transverse flow velocity and the gradual reduction of the non-extensivity parameter towards the Boltzmann-Gibbs limit indicate an increasing level of collectivity and approach to local equilibria in high-multiplicity pp events. Similar collective phenomena, such as long-range azimuthal correlations and strangeness enhancement, have been established experimentally in small systems at the LHC and are generally considered hallmarks of collective dynamics. Nevertheless, our point is that these trends are not exclusive signatures of a genuine QGP phase, and may also appear due to other types of microscopic mechanisms, e.g., color reconnection, multiparton interactions, percolation effects or hadronic rescattering. Therefore, the present results must be considered indicative of strengthened collective dynamics for high multiplicity small systems and not as unambiguous evidence for the formation of QGP. More experimental and theoretical studies are still needed to help unentangle these conflicting interpretations.

In the present work, the analysis is limited to charged pions, which are the most abundant hadronic species in the pp collisions and are dominant in the production of the bulk particle. Owing to their large yields and large contributions from resonance decays, pions are a good proxy for probing the average freeze-out conditions of the system. In comparison, heavier hadrons like kaons and protons have much smaller multiplicities in small systems, and their spectra have a much stronger influence from the role of mass dependence, baryon number conservation, and statistical limitations, which makes it difficult to reliably extract collective flow parameters. While the observation of mass order of flow observables is well known to be expected in the presence of collective dynamics, a statistically robust multi-species analysis in pp collisions requires much larger data samples and is beyond the scope of the present study. Previous experimental and phenomenological investigations have, however, found harder spectra in transverse momenta and flow-like behaviour in kaons and protons in events of high multiplicity pp collisions \cite{identified}, which would suggest qualitatively consistent trends from one hadron species to another. We thus expect that the observed multiplicity dependence of the extracted parameters will also hold for heavier hadron, although differences might be expected due to the above-mentioned different particle properties related to earlier kinetic freeze-out or species-dependent temperature evolution. This limitation of our work is explicitly recognized, and a systematic extension of the present analysis to kaons and protons is identified as an important direction for future work.

It is known that the Tsallis parameters $T$ and $q$ obtained using single-component Tsallis fits may be sensitive to the fitting interval of $p_T$, especially in central heavy-ion collisions, when small $p_T$ windows are used \cite{nath2021centrality}. In our current work, we make a hybrid Hydro+Tsallis (BGBW+Tsallis) formulation in which soft (collective) and hard (non-equilibrium) particle production mechanisms are explicitly separated. The construction allows for a simultaneous and consistent description of the entire measured $p_T$ range, and in this way, it makes the extracted Tsallis parameters less sensitive to the low-$p_T$ regime where collective flow dominates. The same fitting strategy is applied to all the multiplicity classes, and the same $p_T$ coverage is used so that the multiplicity-dependent trends observed do not reflect different fit windows. Although the absolute values of $T$ and $q$ can vary when different intervals of $p_T$ are chosen, the relative trend of these parameters with charged-particle multiplicity, i.e., the rise of $T$ and the fall of $q$, has been found to remain unchanged and physically significant in the chosen framework. An investigation of the dependence of the fit-range of the hybrid Hydro+Tsallis model, similar to the specific study carried out to fit single-component Tsallis in Ref. \cite{nath2021centrality} is a timely and critical issue, which may be the center of future work.

\begin{table*}[!ht]
    \centering
    \caption{The free parameters, $a$ and $b$, and $\chi^2/NDF$ obtained from the fitting procedure of the logarithmic function $y = a \ln(1 + b \frac{dN_{ch}}{d\eta})$. }
    \begin{tabular}{cccccc}
    \hline
    \hline
        Figure & Function & Model & $a$ & $b$ & $\chi^2/NDF$ \\ 
        \hline
    \hline
        Fig. 3(a) & Logarithmic & Tsallis & 0.0396±0.0011 & 8.4974±1.0550 & 2.93E-05 \\ 
        Fig. 3(d) & function & Tsallis & 0.1109±0.0038 & 5.0716±0.6661 & 1.19E-04 \\ 
        \hline
    \hline
    \end{tabular}
\end{table*}
\begin{table*}[!ht]
    \centering
    \caption{The values of free parameters, including slopes $m$ and intercepts $C$ obtained through the fitting via linear function $y=m(\frac{dN_{ch}}{d\eta})+C$ and $y=m\beta_T+C$ for Fig. 4(c).}
    \begin{tabular}{cccccc}
    \hline
    \hline
        Figure & Function & Model & $m$ & $C$ & $\chi^2/NDF$ \\ 
        \hline
    \hline
        Fig. 3(a) & ~ & Hydro+Tsallis & 0.0029±1.34E-4 & 0.1419±0.0015 & 3.21E-05 \\ 
        Fig. 3(b) & ~ & Tsallis & -5.43E-4±1.13E-5 & 1.1826±1.29E-4 & 4.01E-08 \\ 
        Fig. 3(b) & ~ & Hydro+Tsallis & -4.21E-4±2.59E-5 & 1.1846±2.96E-4 & 2.12E-07 \\ 
        Fig. 3(c) & Linear & Tsallis & 93.4564±1.5435 & 102.8953±13.2098 & 0.7484 \\ 
        Fig. 3(c) & function & Hydro+Tsallis & 100.0228±1.2791 & 86.8765±10.8290 & 0.4961 \\ 
        Fig. 3(d) & ~ & Hydro+Tsallis & 0.0080±3.63E-4 & 0.3450±0.0040 & 9.47E-05 \\ 
        Fig. 4(c) & ~ & Hydro+Tsallis & -0.1311±0.0054 & 0.2169±0.0031 & 1.89E-06 \\ 
        Fig. 6(d) & ~ & Hydro+Tsallis & 4.4040±0.03720 & -235.0655±4.2557 & 0.0022 \\ 
        \hline
    \hline
    \end{tabular}
\end{table*}
\begin{table*}
\small
    \centering
    \caption{The values of different free parameters extracted from the fitting procedure of the non-extensive Tsallis-like function, $y=a[1+(q-1)b(dN_{ch}/d\eta))]^{1/(1-q)}$. The model in the third column means the model or function used to fit the experimental $p_T$ spectra data in Fig. 1.}
    \begin{tabular}{cccccccc}
    \hline
    \hline
        Figure & Function & Model & $q$ & $a$ & $b$ & $\chi^2/NDF$ \\ 
            \hline
    \hline
        Fig. 3(b) & \{Non & Tsallis & $q'$=1.1759 & 1.1826±1.30E-4 & 4.6197E-4±9.62E-6 & 4.02E-8 \\ 
        Fig. 3(b) & extensive & Hydro+Tsallis & $q'$=1.1830 & 1.1846±2.96E-4 & 3.57E-4±2.19E-5 & 1.78E-7 \\ 
        Fig. 6(b) & Tsallis & Tsallis  & 1.9041±0.1195 & 5600±2.1E7 & 5695±2.0E7 & 4.62E-4 \\ 
        Fig. 6(b) & function\} & Hydro+Tsallis & 1.7771±0.0605 & 0.1420±0.0062 & 0.2431±0.0250 & 1.72E-5 \\ 
            \hline
    \hline
    \end{tabular}
\end{table*}
\begin{table*}
\small
    \centering
    \caption{The free parameters along with $\chi^2/\text{NDF}$ obtained through the fitting of the sigmoidal function $y=\left[\frac{(y_0-y_{\text{mini}})}{1+\exp{\left(\frac{dN_{\text{ch}}/d\eta - M}{\Delta}\right)}}\right] + y_{\text{mini}}$, where $y$ represents the thermodynamic quantity or response function present along the vertical axis in the plots.}
    \begin{tabular}{cccccccc}
    \hline
    \hline
        Figure & Function & Model & $y_0$ & $y_{mini}$ & $M$ & $\Delta$ & $\chi^2/NDF$ \\
        \hline
    \hline
        Fig. 4(b) & ~ & Hydro+Tsallis & 0.4949±0.0174 & 0.6229±0.0051 & 7.3565±1.1017 & 4.3061±0.8600 & 1.38E-05 \\ 
        Fig. 5(a) & ~ & Hydro+Tsallis & 0.0264±0.0243 & 0.3008±0.0259 & 11.9871±0.6984 & 6.5749±1.4705 & 1.01E-5 \\ 
        Fig. 5(b) & ~ & Hydro+Tsallis & 0.0428±0.0145 & 0.2104±0.0133 & 10.8546±0.6620 & 5.8569±1.2242 & 4.58E-5 \\ 
        Fig. 5(c) & Sigmoidal & Hydro+Tsallis & 0.4205±0.1462 & 1.8385±0.1234 & 10.5722±0.7737 & 5.9726±1.4151 & 4.23E-4 \\ 
        Fig. 5(d) & ~ & Hydro+Tsallis & 0.0117±0.0063 & 0.0948±0.0081 & 11.9654±0.7430 & 5.9210±1.2830 & 2.98E-5 \\  
        Fig. 5(e) & ~ & Hydro+Tsallis & 1.8501±0.3622 & 5.3718±0.3946 & 11.2123±0.9262 & 3.9868±1.2726 & 0.0073 \\ 
        Fig. 5(f) & ~ & Hydro+Tsallis & 0.3264±0.0018 & 0.3312±3.23E-4 & 5.3693±3.8739 & 5.8280±2.2851 & 3.39E-8 \\ 
        \hline
    \hline
    \end{tabular}
\end{table*}
\begin{table*}[!ht]
    \centering
    \caption{The free parameters, $a$ and $b$, and $\chi^2/NDF$ obtained from the fitting procedure of the power law function $y=ax^b$.}
    \begin{tabular}{cccccc}
    \hline
    \hline
        Figure & Function & Model & $a$ & $b$ & $\chi^2/NDF$ \\ 
        \hline
    \hline
        Fig. 5(a) &~& Tsallis & 0.0192±0.0020 & 0.8741±0.0414 & 7.61E-4 \\ 
        Fig. 5(b) &~& Tsallis & 0.0241±0.0020 & 0.7074±0.0336 & 4.49-E-4 \\
        Fig. 5(c) & Power Law & Tsallis & 0.2409±0.0202 & 0.6559±0.0344 & 0.0043 \\
        Fig. 5(d) & function & Tsallis & 0.0061±6.45E-4 & 0.8832±0.0421 & 2.56E-4 \\
        Fig. 5(e) &~& Tsallis & 0.7432±0.0612 & 0.6490±0.0337 & 0.0126 \\
        Fig. 6(d) &~& Tsallis & -424.340±19.299 & -0.3671±0.0220 & 0.4387 \\
        \hline
    \hline
    \end{tabular}
\end{table*}
\begin{table*}[!ht]
    \centering
    \caption{The values of the free parameter and $\chi^2/NDF$ extracted from the fitting of the exponential decay model $y = y_0 \exp(-\alpha \frac{dN_{ch}}{d\eta}) + y_{\text{min}}$.}
    \begin{tabular}{ccccccc}
    \hline
    \hline
        Figure & Function & Model & $y_0$ & $\alpha$ & $y_{mini}$ & $\chi^2/NDF$ \\ 
        \hline
    \hline
        Fig. 6(a) & ~ & Tsallis & 44.4657±7.1902 & 0.3842±0.0502 & 5.3147±0.3991 & 0.1004 \\ 
        Fig. 6(a) & \{Exponential & Hydro+Tsallis & 10.8682±0.3620 & 0.0994±0.0122 & 3.3424±0.4757 & 0.0065 \\ 
        Fig. 6(b) & decay & Tsallis & 0.3980±0.0664 & 0.4142±0.0498 & 0.0232±0.0026 & 0.001 \\ 
        Fig. 6(b) & function\} & Hydro+Tsallis & 0.1057±0.0017 & 0.1678±0.0048 & 0.0159±0.0504 & 1.14E-05 \\ 
        Fig. 6(c) & ~ & Tsallis & 325.9560±58.9655 & 0.4520±0.0564 & 20.2940±1.9658 & 0.7588 \\ 
        Fig. 6(c) & ~ & Hydro+Tsallis & 57.2494±1.9293 & 0.1129±0.0127 & 11.3042±2.0057 & 0.0464 \\ 
        \hline
    \hline
    \end{tabular}
\end{table*}

\section{Conclusion}
Although QGP-like behaviour has traditionally been associated with heavy-ion collisions, recent studies have identified signatures of collective phenomena (such as azimuthal anisotropies and long-range correlations) even in pp collisions at the LHC, prompting further investigation into the system's thermodynamic and collective properties in smaller collision systems.
%Although QGP-like behaviour has traditionally been associated with heavy-ion collisions, recent studies have identified signatures such as collective behaviour (azimuthal anisotropies and long-range correlations) even in pp collisions at the LHC, prompting further investigation into the system's thermodynamics and hydrodynamics produced in the smaller collision systems. 
The \( p_T \) distributions of pions (\( \pi^+ + \pi^- \)) up to \( p_T = 20 \) GeV/c produced in ten distinct multiplicity classes of pp collisions at \( \sqrt{s} = 7 \) TeV were analyzed. Two models were employed for fitting the \( p_T \) distributions: the Tsallis-Pareto type function and the combined Boltzmann-Gibbs Blast Wave and Tsallis-Pareto type model (denoted as Hydro+Tsallis). The fitting process was conducted using the minimum \( \chi^2 \) method, and the Hydro+Tsallis model provided a more accurate representation of the \( p_T \) spectra compared to the Tsallis-Pareto model.
Various parameters were extracted through the fitting procedure, including the Tsallis temperature, non-extensivity parameter, normalization constant, kinetic freeze-out temperature, transverse flow velocity, and mean transverse momentum. The Tsallis-Pareto model yielded values for \( T \), \( q \), \( N_0 \), and \( \langle p_T \rangle \), while the Hydro+Tsallis model provided \( T_0 \), \( \beta_T \), \( T \), \( q \), \( N_0 \), and \( \langle p_T \rangle \). Using the extracted values of \( T \) and \( q \), a range of thermodynamic quantities at the freeze-out stage were calculated, such as energy density, particle density, entropy density, pressure, and additional thermodynamic response functions including specific heat at constant volume, squared speed of sound, mean free path, Knudsen number, isothermal compressibility, and expansion coefficient.
The study revealed that parameters such as \( T \), \( \langle p_T \rangle \), \( N_0 \), \( \epsilon \), and \( n \) increased with rising charged particle multiplicity density (\( dN_{ch}/d\eta \)), indicating higher energy transfer in events with greater \( dN_{ch}/d\eta \), leading to increased energy density, particle density, and system excitation. The entropy density (\( s \)) also showed an upward trend with increasing \( dN_{ch}/d\eta \), suggesting the production of a highly disordered system in high-multiplicity events. %Additionally, pressure \( P \), flow velocity \( \beta_T \), and expansion coefficient $\alpha$ rose with increasing \( dN_{ch}/d\eta \), possibly indicative of a highly compressed, collective, and hydrodynamic-like system producing greater pressure and enhanced flow effects.
Additionally, pressure \( P \), flow velocity \( \beta_T \), and expansion coefficient $\alpha$ rose with increasing \( dN_{ch}/d\eta \), possibly indicative of enhanced collective phenomena reminiscent of those observed in larger collision systems.

Specific heat at constant volume (\( C_V \)) and squared speed of sound (\( c_s^2 \)) exhibited growth with higher \( dN_{ch}/d\eta \), implying the possible formation of a stiffer system capable of storing thermal energy and supporting compressional waves. In contrast, \( T_0 \) demonstrated an inverse relationship with rising \( dN_{ch}/d\eta \), suggesting the creation of a shorter-lived system at higher multiplicity. Furthermore, the non-extensivity parameter \( q \) decreased as \( dN_{ch}/d\eta \) increased, possibly indicating the development of a more thermally equilibrated, collective, or hydrodynamic system in high-multiplicity events.
Finally, the mean free path (\( \lambda \)), Knudsen number (\( K_n \)), and isothermal compressibility (\( \kappa_T \)) were found to decrease as \( dN_{ch}/d\eta \) increased, suggesting the emergence of stronger collective behavior in high-multiplicity events compared to low-multiplicity ones.
%Finally, the mean free path (\( \lambda \)), Knudsen number (\( K_n \)), and isothermal compressibility (\( \kappa_T \)) were found to decrease as \( dN_{ch}/d\eta \) increased, which might suggest the formation of a more liquid-like system in high-multiplicity events compared to low-multiplicity ones. 
Correlations among the extracted parameters were also investigated, adding further insight into the behavior of the system under study.\\

\noindent {\bf Author Contributions}:
\textbf{Murad Badshah}: Conceptualization; Methodology; Formal analysis; Investigation; Data curation; Software; Visualization; Writing -- original draft; Writing -- review \& editing.
\textbf{Haifa I. Alrebdi}: Conceptualization; Methodology; Validation; Resources; Funding acquisition; Writing -- review \& editing; Supervision.
\textbf{Muhammad Waqas}: Formal analysis; Software; Data curation; Visualization; Writing -- review \& editing.
\textbf{Hadiqa Qadir}: Investigation; Data curation; Validation; Visualization; Writing -- review \& editing.
\textbf{Muhammad Ajaz}: Conceptualization; Methodology; Supervision; Project administration; Resources; Funding acquisition; Writing -- original draft; Writing -- review \& editing.

\medskip
All authors have read and approved the final version of the manuscript and agree to be accountable for all aspects of the work.

\noindent {\bf Data availability}
The data used to support the findings of this study are included within the article and are cited at relevant places within the text as references.
\\

\noindent {\bf Compliance with Ethical Standards}
The authors declare that they comply with ethical standards regarding the content of this paper.
\\

\noindent {\bf Acknowledgments}
This research work was supported by Princess Nourah bint Abdulrahman University Researchers Supporting Project number (PNURSP2026R106), Princess Nourah bint Abdulrahman University, Riyadh, Saudi Arabia. The authors would like to thank Professor Dr. Li Yi, Shandong University, China, for her insightful suggestions on the improvement of the current work.
The authors would also like to thank Abdul Wali Khan University Mardan, Pakistan, for providing the opportunity to conduct this research. 
\\

\noindent {\bf The authors declare that they have no known competing financial interests or personal relationships that could have appeared to influence the work reported in this paper.}
\bibliographystyle{aip}
\bibliography{references}
\end{multicols}
\end{document}